\documentclass[12pt,reqno]{article}

\usepackage{amsmath}
\usepackage{amssymb}
\usepackage{amsthm}
\usepackage{bm}
\usepackage{booktabs}
\usepackage[bf,footnotesize]{caption}
\usepackage{changepage}
\usepackage{color}
\usepackage{eucal}
\usepackage{fullpage}
\usepackage{graphicx}
\usepackage{hhline}
\usepackage{lineno}
\usepackage{makecell}
\usepackage[sc]{mathpazo}
\usepackage{mathrsfs}
\usepackage{mathtools}
\usepackage{multirow}
\usepackage[numbers,sort&compress,square]{natbib}
\usepackage{setspace}
\usepackage{stmaryrd}
\usepackage[all,cmtip]{xy}

\newcommand*\patchAmsMathEnvironmentForLineno[1]{%
	\expandafter\let\csname old#1\expandafter\endcsname\csname #1\endcsname
	\expandafter\let\csname oldend#1\expandafter\endcsname\csname end#1\endcsname
	\renewenvironment{#1}%
	{\linenomath\csname old#1\endcsname}%
	{\csname oldend#1\endcsname\endlinenomath}}%
	\newcommand*\patchBothAmsMathEnvironmentsForLineno[1]{%
	\patchAmsMathEnvironmentForLineno{#1}%
	\patchAmsMathEnvironmentForLineno{#1*}}%
	\AtBeginDocument{%
	\patchBothAmsMathEnvironmentsForLineno{equation}%
	\patchBothAmsMathEnvironmentsForLineno{align}%
	\patchBothAmsMathEnvironmentsForLineno{flalign}%
	\patchBothAmsMathEnvironmentsForLineno{alignat}%
	\patchBothAmsMathEnvironmentsForLineno{gather}%
	\patchBothAmsMathEnvironmentsForLineno{multline}%
}

\newcommand{\eq}[1]{\textbf{Eq.~\ref{eq:#1}}}
\newcommand{\fig}[1]{\textbf{Fig.~\ref{fig:#1}}}
\newcommand{\vid}[1]{\textbf{Vid.~\ref{vid:#1}}}

\newcommand{\sect}[1]{Section~\ref{sec:#1}}

\newcommand{\p}{\mathbf{p}}
\newcommand{\q}{\mathbf{q}}

\title{\begin{center} \bfseries\singlespacing
Selfish optimization and collective learning in populations
\end{center}}
\author{\parbox[c]{16cm}{\onehalfspacing \normalsize \centering ~\\[-0.4cm] Alex McAvoy$^{1,2}$ ,  Yoichiro Mori$^{1,2,3}$, and Joshua B. Plotkin$^{1,2,3}$ \\ \quad\\ \footnotesize
$^{1}$Department of Mathematics, University of Pennsylvania, Philadelphia, PA 19104, USA \\
$^{2}$Center for Mathematical Biology, University of Pennsylvania, Philadelphia, PA 19104, USA \\
$^{3}$Department of Biology, University of Pennsylvania, Philadelphia, PA 19104, USA\\[0.2cm]}
\date{}
}

\begin{document}

\allowdisplaybreaks

\maketitle

\begin{abstract}
A selfish learner seeks to maximize their own success, disregarding others. When success is measured as payoff in a game played against another learner, mutual selfishness typically fails to produce the optimal outcome for a pair of individuals. However, learners often operate in populations, and each learner may have a limited duration of interaction with any other individual. Here, we compare selfish learning in stable pairs to selfish learning with stochastic encounters in a population. We study gradient-based optimization in repeated games like the prisoner's dilemma, which feature multiple Nash equilibria, many of which are suboptimal. We find that myopic, selfish learning, when distributed in a population via ephemeral encounters, can reverse the dynamics that occur in stable pairs. In particular, when there is flexibility in partner choice, selfish learning in large populations can produce optimal payoffs in repeated social dilemmas. This result holds for the entire population, not just for a small subset of individuals. Furthermore, as the population size grows, the timescale to reach the optimal population payoff remains finite in the number of learning steps per individual. While it is not universally true that interacting with many partners in a population improves outcomes, this form of collective learning achieves optimality for several important classes of social dilemmas. We conclude that na\"{i}ve learning can be surprisingly effective in populations of individuals navigating conflicts of interest.
\end{abstract}

\section{Introduction}
Population-based methods of optimization and learning have received considerable attention over the years. These methods are often inspired by natural selection and attribute ``fitness'' to candidate solutions, with higher fitness corresponding to better solutions, for a given problem at hand. Genetic algorithms are classical examples, which encode solutions using an alphabet (e.g.~as binary strings) and are amenable to crossover, mutation, and fitness-based reproduction \citep{goldberg:book:1989,schmitt:TCS:2001}. Applications are broad, although these algorithms are especially common in problems with an overwhelming number of candidate solutions, such as the traveling salesman problem \citep{larranaga:AIR:1999} or the search for appropriate weights and topologies in artificial neural networks (``neuroevolution'') \citep{stanley:EC:2002}.

In some classes of population-based optimization, the population itself is structured according to the search space, with individuals corresponding to locations in parameter space. Particle swarm optimization, for example, involves solutions moving through parameter space as ``particles'' with some velocity \citep{kennedy:ICNN:1995}. The velocity is influenced by the best solutions found to date, both by the individual particle and by the population as a whole (or at least locally within a neighborhood of the particle). One reason why this approach is successful is that it promotes exploration, as particles tend to overshoot local optima due to momentum \citep{eberhart:MHS:1995}.

In such metaheuristics, populations have proven remarkably effective in finding good solutions to computationally hard problems. But these problems are frequently characterized by static fitness landscapes, meaning the presence of other agents does not affect the viability of a solution found by a particular agent. In contrast, the goal of multi-agent reinforcement learning is to optimize an agent's success or fitness in the presence of other agents, who themselves might be carrying out a similar kind of optimization \citep{panait:AAMAS:2005,hoen:LAMAS:2006}. A standard model for this kind of problem is a stochastic game, which involves repeated interactions among a group of individuals in some (possibly changing) environment \citep{shapley:PNAS:1953,littman:MLP:1994}.

Even in a simple setting without a dynamic environment, multi-agent learning is far from completely understood. A simple example is a repeated game between two individuals. Folk theorems \citep{fudenberg:E:1986,barlo:JET:2009} imply that such games often have a rich variety of equilibria, which is most striking when the strategic interaction is ``mixed,'' meaning it is neither purely cooperative (aligned incentives) nor purely competitive (zero-sum) \citep{zhang:arXiv:2019}. The traditional example is the repeated prisoner's dilemma, which has only detrimental equilibria when the game horizon is short, but many different (partly) cooperative equilibria when the horizon is longer \citep{nowak:Nature:2006}. The ability of long interaction horizons to reinforce prosocial behaviors has been referred to as the ``shadow of the future'' \citep{axelrod:BB:1984}, and it is one of the central tenets of the theory of direct reciprocity \citep{trivers:TQRB:1971,hilbe:NHB:2018}.

Given the rich set of equilibria in repeated games, there remains the question of whether two self-concerned learners will jointly arrive at good outcomes. The answer, perhaps unsurprisingly, is no--at least not reliably. Mutual selfishness means that agents often cannot overcome the temptation to exploit the opponent, eroding prosocial actions and leading to detrimental outcomes for both individuals, despite the existence of cooperative equilibria. A recent study \citep{foerster:AAMAS:2018} proposes a clever way to overcome this drawback of selfish (``na\"{i}ve'') optimization by attempting to shape the co-player's future optimization problem. This learning rule, dubbed ``learning with opponent-learning awareness,'' can better align incentives of two players by utilizing a short look-ahead into the co-player's learning, provided the co-player's learning rule is known. However, as in much of the work on multi-agent learning, this learning rule is implemented under the assumption that learners interact in stable pairs, which means that an agent spends a significant amount of time optimizing against a fixed opponent (who is also learning).

In a population of learners, however, there is no guarantee that any given encounter will be long enough to resemble stable learning. An agent might enter the world with little training, equipped only with a learning rule and a rough idea of how to engage with others. Encounters might be ephemeral, requiring a learner to efficiently use limited information to update its style of play, which is then brought forth into future engagements with different partners. Moreover, the goal in designing a learning algorithm may not be to produce a single agent who performs well, but rather an entire population that plays well with one another. Automated vehicles are agents like this, which are extensively trained prior to ever interacting with another car or person in the real world but must nonetheless sense and adjust to previously unseen encounters with other vehicles and pedestrians \citep{hancock:PNAS:2019}. The goal of training for automated vehicles is assuredly not to produce a single vehicle capable of optimal performance, disregarding the performance of other vehicles -- but, rather, to produce an entire population of agents who all perform well. Likewise human beings, whose capabilities are often targeted as goals of and inspiration for artificial intelligence, also learn from many different partners throughout life, and they may benefit from a population that collectively performs well.

In this study, we consider a blended population-based framework, with learning in place of strategy transmission and transient encounters in place of stable pairs. We disregard ``vertical'' transmission and consider how social encounters influence the behaviors of individuals within a fixed population of learners. The population has no central planner or controller to enforce sophisticated social preferences; there are only random encounters and selfish (greedy) strategy updates. Unlike population-based optimization techniques on static landscapes, in which an outcome might be considered successful even if only a small portion of the population finds an optimal solution (including genetic algorithms and some swarm intelligence methods \citep{zielinski:I:2007}), our goal here is not to produce a small subset of fit individuals in a population. Rather, we seek to understand how social learning in dynamic optimization landscapes affects members of a population collectively.

Even when selfish learning in stable pairs leads to poor outcomes, as in the repeated prisoner's dilemma, we will show that selfish learning in a population with ephemeral partnerships can collectively attain the maximum possible average payoff. Moreover, the timescale to attain such an optimum does not necessarily slow down for any particular individual as the population size (and thus the number of potential partners) grows. We conclude that sociable, selfish optimizers, who learn little from any one encounter, can nonetheless be quite effective and efficient promoters of prosperity and collective learning within a population. In particular, the ability of agents to have a variety of partners within a population can drastically improve the efficacy of selfish learning in the presence of conflicts of interest.

\section{Model}
The players in our model learn how to behave in repeated games. Each learning step consists of several rounds of a repeated game, followed by a strategy update step. When player $X$ interacts with player $Y$, they play a repeated game with a stochastic number of rounds determined by a continuation probability, $\lambda <1$. Each round is of a one-shot (stage) game with actions $A$ and $B$ and payoffs
\begin{align}
\bordermatrix{%
& A & B \cr
A &\ a_{AA} & \ a_{AB} \cr
B &\ a_{BA} & \ a_{BB} \cr
}\ . \label{eq:payoff_matrix}
\end{align}
In \eq{payoff_matrix}, the values indicate what payoff the row player receives when facing the column player. After each round, the game continues with probability $\lambda$ and ends with probability $1-\lambda$. Here, we care about repeated games with sufficiently long time horizons, meaning $\lambda$ is close to $1$. We take $\lambda =0.9999<1$ for an average of $10^{4}$ rounds in the repeated game, both because all interactions are finite (even if long) and because the limit $\lambda\rightarrow 1$ can introduce technical issues (see \sect{sm_payoffs}).

A memory-one strategy for $X$ in this repeated game consists of an initial probability of playing $A$, $p_{0}\in\left[0,1\right]$, together with a four-tuple of conditional probabilities, $\left(p_{AA},p_{AB},p_{BA},p_{BB}\right)\in\left[0,1\right]^{4}$, where $p_{xy}$ indicates $X$'s probability of playing $A$ when, in the previous round, $X$ played $x$ and $Y$ played $y$. We denote a player's memory-one strategy, $\left(p_{0},\left(p_{AA},p_{AB},p_{BA},p_{BB}\right)\right)$, by $\p\in\left[0,1\right]^{5}$ to simplify notation. We denote by $\pi\left(\p ,\q\right)$ the payoff to an individual using $\p$ against an opponent using $\q$. How this payoff is calculated, together with its functional form, is detailed in \sect{sm_payoffs}.

A learning rule consists of \emph{(i)} an initial distribution over strategies, representing a prior policy before any learning takes place, and \emph{(ii)} an update rule for how to adjust this policy after an encounter. Since we consider memory-one strategies here, the initial policy is chosen from a distribution on $\p\in\left[0,1\right]^{5}$. In practice, we consider two kinds of initial distributions. The primary distribution we analyze chooses each of the five coordinates independently from an arcsine ($\textrm{Beta}\left(1/2,1/2\right)$) distribution. For each coordinate, this distribution explores the corners of $\left[0,1\right]$ better than a uniform distribution does, and it is well-established in models of repeated interactions \citep{nowak:Nature:1993}. However, a uniform distribution is also a natural choice, and so we also consider examples in which each coordinate is sampled uniformly (and independently) in $\left[0,1\right]$.

Suppose that $X$ and $Y$ are selfish learners paired to interact. After the repeated game is completed, the players perform a gradient-based strategy update step, with both players updating simultaneously. If $X$ has strategy $\p$ and $Y$ has strategy $\q$ prior to this update, then their respective strategies after one gradient step are
\begin{subequations}\label{eq:PGA}
\begin{align}
\p ' &= \textrm{proj}\left(\p + \eta \nabla_{\p} \pi \left(\p ,\q\right)\right) ; \\
\q ' &= \textrm{proj}\left(\q + \eta \nabla_{\q} \pi \left(\q ,\p\right)\right) ,
\end{align}
\end{subequations}
where $\eta$ is a small learning rate and $\textrm{proj}$ denotes the projection operator onto the cube $\left[0,1\right]^{5}$, which ensures that strategies remain in $\left[0,1\right]^{5}$. Note that even if two learners both update strategies using \eq{PGA}, we consider their learning rules to be different if they have different initial distributions.

\begin{figure}
\centering
\includegraphics[width=0.8\textwidth]{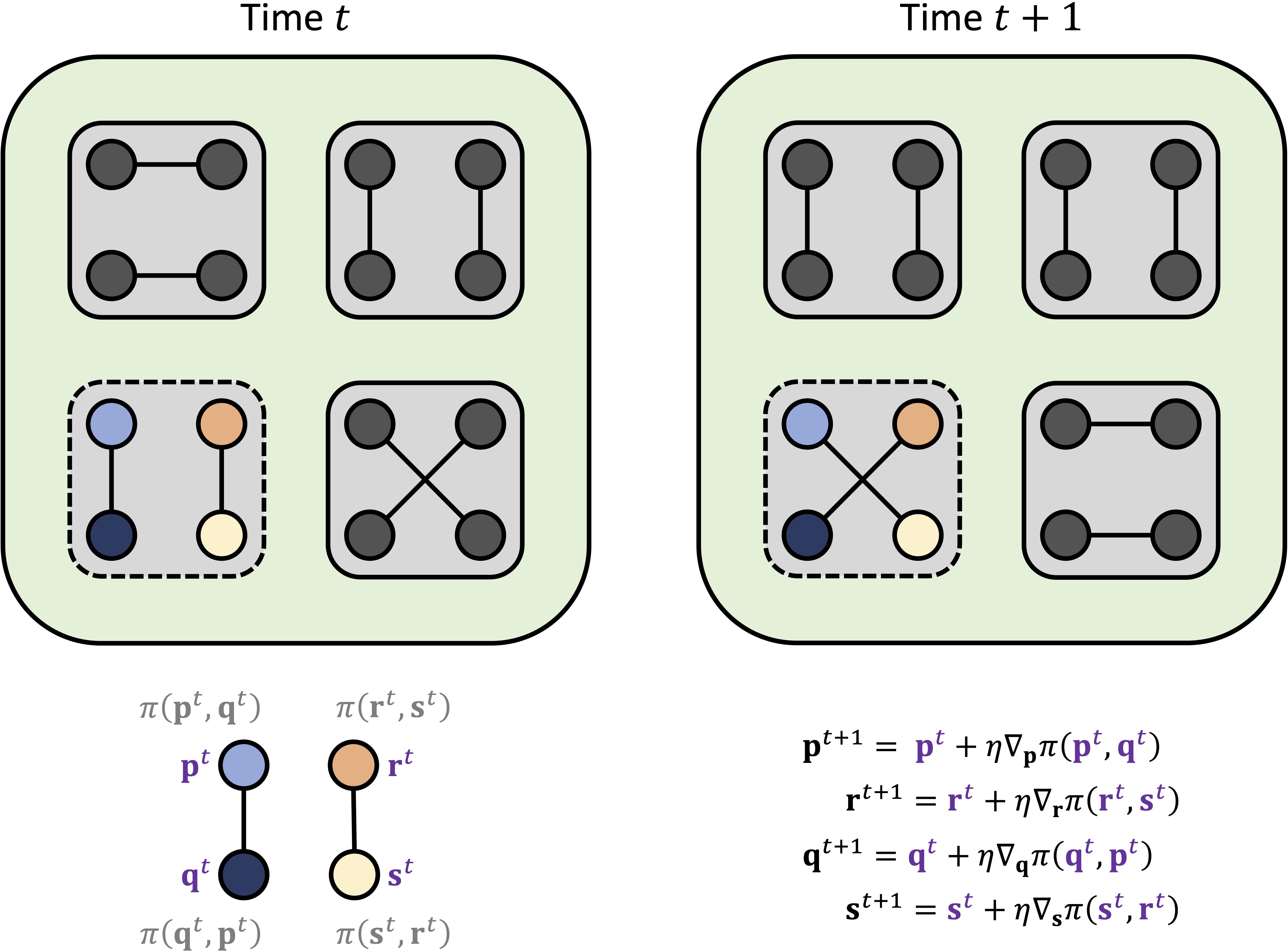}
\caption{\textbf{Learning in isolated sub-populations.} The figure depicts a population of $N$ total individuals arranged in $d$ sub-populations, each of size $n$ ($d=n=4$ and $N=dn=16$ shown here). Individuals are paired randomly within each sub-population, receive scores from their interactions (based on $\pi$), and perform gradient-based strategy updates. The updated strategies are then brought forth into subsequent encounters, which may involve new partnerships (depicted here, for example, in the dotted sub-population). In two extreme cases, individuals interact with the same partner in every time step (stable pairs, $n=2$) or with potentially any other member of the entire population (well-mixed population, $d=1$). To simplify notation, the projection operator used in the gradient update is omitted.\label{fig:population_cartoon}}
\end{figure}

To describe how learners are paired with partners, we consider a population of $d$ disjoint sub-populations, each of size $n$, where $n$ is even. The total population size is thus $N=dn$. When $n=2$, all pairings are between the same two individuals, and the model reduces to a population whose individuals learn in stable pairs. In each time step, individuals are randomly paired within each sub-population to interact and update their strategies via \eq{PGA}. In the next time step, players are again paired randomly within each sub-population, bringing their new strategies from the previous time step, and the process repeats. Thus, there are two timescales in our model: one timescale for repeated actions in a one-shot game, and one timescale for the strategy updates (``learning'') following the repeated game. To make the distinction clear, we refer to the repeated one-shot games as ``interactions'' and the combination of pairing, a repeated game, and a strategy update as an ``encounter.'' This population-based learning process is depicted schematically in \fig{population_cartoon}.

For fixed population size $N=dn$, we are interested in comparing $n=2$ to $n>2$ to understand the effects of distributed learning in a population compared to the standard case of stable pairings. When $n>2$, an individual's partner can change across encounters. An alternative approach would be to set $d=1$ and study the effects of varying $N$ ($=n$). Apart from one example, we have avoided this approach because it involves the comparison of populations of different size. We care about the mean payoff of the population, and changing $N$ while keeping $d$ fixed would involve averaging over a different number of individuals. In contrast, with $N$ fixed and $n$ varying, the population always starts off with the same distribution of mean payoff prior to learning, regardless of $n$, which permits a more straightforward understanding of the effects of randomness in partner choice on learning.

At any time, everyone has a strategy for the repeated game. When they are paired to play repeated games based on these strategies, the average payoff in each pair satisfies
\begin{align}
\min_{x,y\in\left\{A,B\right\}}\left\{\frac{a_{xy}+a_{yx}}{2}\right\} \leqslant \frac{\pi\left(\p ,\q\right) +\pi\left(\q ,\p\right)}{2} \leqslant \max_{x,y\in\left\{A,B\right\}}\left\{\frac{a_{xy}+a_{yx}}{2}\right\} . \label{eq:payoff_interval}
\end{align}
As a result, the mean payoff in the population also lies in this interval. The main question we study is how increasing $n$ (with $N$ fixed) affects this mean population payoff in the long run.

\section{Results}
We analyze our model of collective learning using two complementary approaches, ``Lagrangian'' and ``Eulerian,'' borrowing these terms from fluid dynamics \citep{batchelor:CUP:2000}. In the Lagrangian approach, we track the behaviors of individual learners in finite populations. Here, we are not concerned with quantitative effects of learning as a function of population size; instead, we wish to explore what the learning process looks like when the population is reasonably large. In particular, we will use the Lagrangian perspective to show that collective learning can produce optimal or near-optimal solutions (e.g. in prisoner's dilemma games) in populations of even moderate size. In the Eulerian approach, we consider the density of regions within the strategy space in an infinite population. This approach is used to take the large-population limit, and it produces results that reaffirm the qualitative findings obtained using the Lagrangian approach. The Eulerian approach will also reveal an important result on the timescale of learning in prisoner's dilemma games: convergence to the optimal population outcome requires only finitely many learning steps per individual, even when the population is arbitrarily large.

\subsection{Lagrangian perspective: tracking individual agents}
Within the space of all payoff matrices in the form of \eq{payoff_matrix}, there are a several distinguished classes of games. The most common example is a prisoner's dilemma, which satisfies $a_{BA}>a_{AA}>a_{BB}>a_{AB}$. This payoff ranking implies that if the game is played once, $B$ (called ``defection'') is the unique Nash equilibrium because an individual always improves its payoff by switching from $A$ to $B$. Both players would prefer the payoff for $A$ (called ``cooperation'') against $A$ to the payoff for $B$ against $B$. When the game is repeated, however, multiple equilibria exist and may bring different long-term payoffs, ranging from the small (defecting-type equilibria) to the large (cooperative equilibria).

\begin{figure}
\centering
\includegraphics[width=0.9\textwidth]{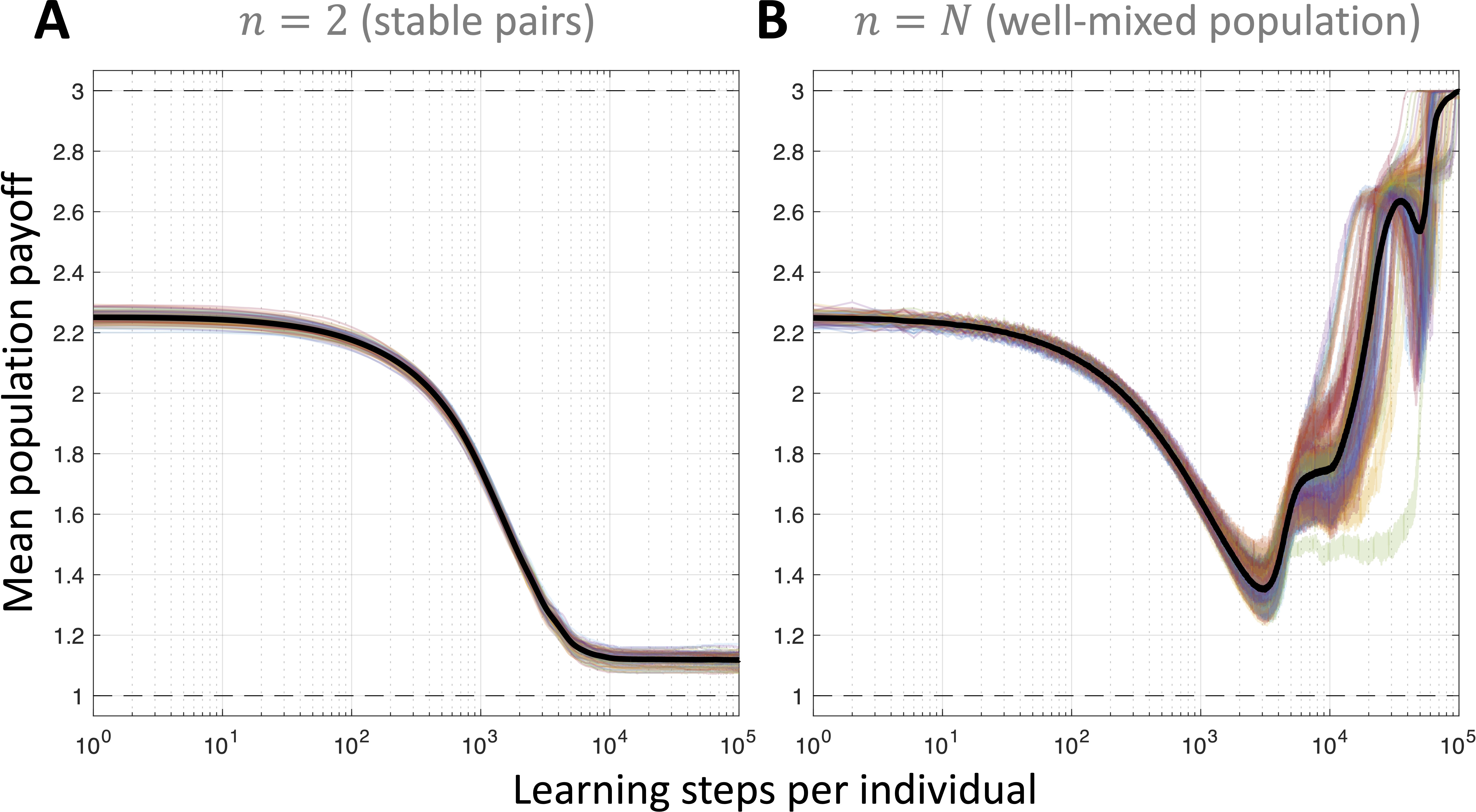}
\caption{\textbf{Pairwise versus collective learning in a repeated prisoner's dilemma.} In both panels, each trajectory (colored) represents the mean payoff of a population of $N=1{,}000$ players as learning unfolds. In \textbf{A}, the population is sorted once into $d=N/2$ sub-populations, each of size two, and learning takes place only within these sub-populations. Initial strategies are chosen randomly from an arcsine distribution. In \textbf{B}, there is only one sub-population, and random pairing within the entire population occurs prior to every learning step. The curves are generated by calculating the payoffs of paired partners at each step using their current strategies, and then taking the average payoff of the population. Unlike the case $d=N/2$, in which all individuals interact with a fixed learning partner throughout the process, in \textbf{B} the outcome for the population is much better. Following an equilibration period in which the mean population payoff declines, the mean payoff begins to increase (non-monotonically) until it hits the maximal social value of $a_{AA}$ ($=3$). The mean over all $100$ runs is shown in black, on top of each individual run (colored). In both panels, the game parameters are $\left(a_{AA},a_{AB},a_{BA},a_{BB}\right) = \left(3,0,5,1\right)$, the continuation probability is $\lambda =0.9999$, and the learning rate is $\eta =10^{-3}$. The dashed lines indicate the minimum and maximum possible average payoffs, respectively (see \eq{payoff_interval}).\label{fig:standardPD}}
\end{figure}

Selfish learners are not able to reliably find cooperative equilibria when they interact in stable pairs. \fig{standardPD}\textbf{A} shows learning trajectories for a population of stable pairs ($n=2$) when each pair starts with random memory-one strategies prior to optimization. These trajectories consistently lead the population to a low payoff, close to the minimum of $a_{BB}$. In contrast, when learning is distributed throughout a population via ephemeral, random encounters, the mean population payoff eventually converges to the maximum possible, $a_{AA}$ (\fig{standardPD}\textbf{B}). At this point, all individuals in the population have learned to cooperate in the repeated game, even though each one is seeking only to maximize its own payoff. \fig{standardPD_pairs} gives another depiction of \fig{standardPD}\textbf{A}, showing the mean payoff trajectory for each pair. A small number of pairs do converge to optimal payoffs, which is reflected by the solid black curve in \fig{standardPD}\textbf{A} converging to a value slightly above $1$.

Although \fig{standardPD} depicts two extremes, we note that the subpopulation size, $n$, should be sufficiently large to achieve optimal payoffs reliably. Small values of $n$ greater than $2$ need not lead to optimal payoffs for the population (see also \fig{alternatingPD} below). However, there does not appear to be a simple threshold for $n$ and $N$ that guarantees convergence of mean population payoffs to optimal or near-optimal values. Such values are evidently sensitive to the game parameters.

We also consider the prisoner's dilemma when each coordinate of the initial strategy is chosen uniformly at random from $\left[0,1\right]$. \fig{standardPD_uniform} illustrates qualitatively similar results for this initial condition: the population still benefits from flexibility in partner choice. Furthermore, although we are concerned mainly with agents who have access to the exact gradients, we also consider an alternative optimization procedure based on random search \citep{solis:MOR:1981}. Instead of calculating gradients, an agent simply samples a nearby strategy and tests this against the partner. If it improves the agent's payoff, then it is adopted; otherwise, the current strategy is retained. \fig{standardPD_RS} shows, once again, that even for this update rule, collective learning with ephemeral partners still produces excellent outcomes for a population in this repeated prisoner's dilemma.

While it is remarkable that such an outcome can be achieved via selfishness alone, simply by swapping partners sufficiently often, a natural question is whether optimal payoffs will also be attained in different forms of prisoner's dilemmas, when mutual cooperation is inefficient. In particular, when $a_{AA}<\left(a_{AB}+a_{BA}\right) /2$, players fare better with a policy of alternation rather than mutual cooperation--that is, for one player to cooperate in even rounds and defect in odd rounds, and for the other player to cooperate in odd rounds and defect in even rounds. \fig{alternatingPD} shows how increasing the size of the sub-populations (while keeping $N$ fixed) in this kind of prisoner's dilemma influences the mean population payoff. When $n=2$ (stable pairs), most runs converge to defection. As $n$ grows, the population finds cooperative strategies, and for yet larger $n$ the population discovers policies of alternation, which produce the maximum possible mean payoff. As in \fig{standardPD}, throughout the learning process the population initially experiences a decline in payoffs before reaching a trough, beyond which the mean payoff rises until it reaches $\left(a_{AB}+a_{BA}\right) /2$.

\begin{figure}
\centering
\includegraphics[width=\textwidth]{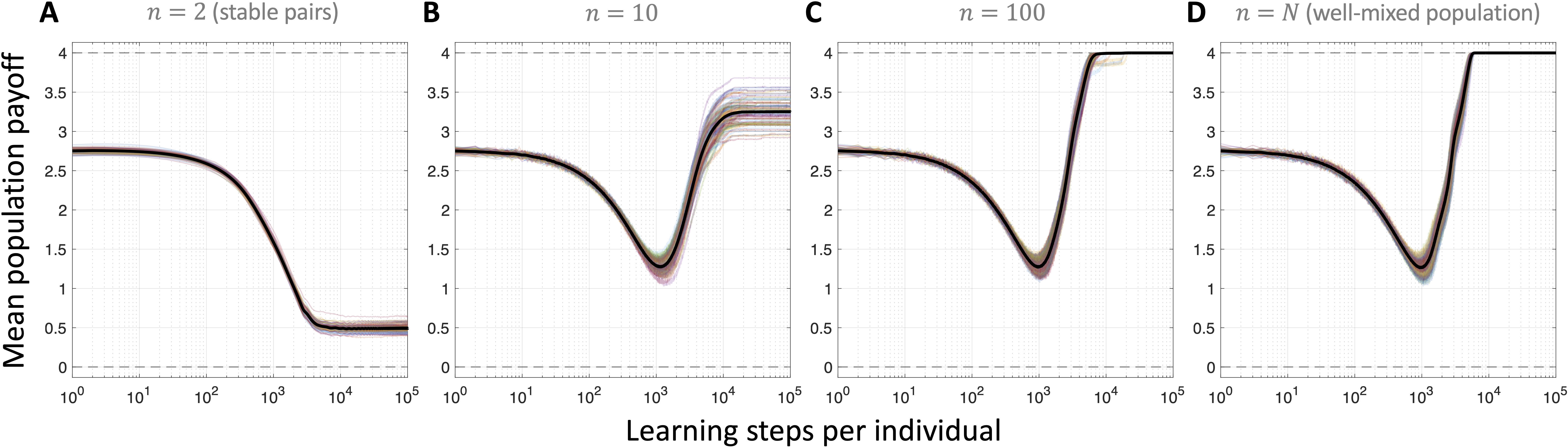}
\caption{\textbf{Collective learning in an alternating prisoner's dilemma.} All four panels depict a repeated prisoner's dilemma with $\left(a_{AB}+a_{BA}\right) /2>a_{AA}$, which means that mutual cooperation is no longer optimal. Rather, a policy of strictly alternating cooperation and defection, with $X$ cooperating in even rounds only and $Y$ cooperating in odd rounds only, is the best social outcome. Although sorting individuals into $d=N/2$ sub-populations of stable pairs still results in bad outcomes (primarily defection), as the population becomes more well-mixed, the optimal outcome of $\left(a_{AB}+a_{BA}\right) /2$ ($=4$) is achieved. Thus, this collective learning process is not simply discovering how to cooperate mutually--it can also find superior outcomes in interactions for which more complex coordination is required. In all panels, the population size is $N=1{,}000$, the number of runs is $100$, the game parameters are $\left(a_{AA},a_{AB},a_{BA},a_{BB}\right) = \left(3,-1,9,0\right)$, the continuation probability is $\lambda =0.9999$, and the learning rate is $\eta =10^{-3}$.\label{fig:alternatingPD}}
\end{figure}

It is also notable in this example that, for fixed $N$, increasing group size $n$ does not necessarily lead to a longer timescale for achieving the optimal population payoff. Increasing $n$ means more randomness in partner selection, and so one might expect slower convergence to a final outcome. And yet, for this repeated prisoner's dilemma, a larger group actually hastens the ascent to the maximal mean payoff.

Both \fig{standardPD} and \fig{alternatingPD} depict prisoner's dilemmas and suggest that \emph{(i)} populations of randomly-interacting selfish learners can attain qualitatively better outcomes than stable selfish learning pairs and \emph{(ii)} for fixed $N$, the mean population payoff is a non-decreasing function of $n$. Given this finding in strong social dilemmas, it is natural to ask whether different behavior is observed in other classes of interactions. A weaker social dilemma \citep{hauert:IJBC:2002} called the snowdrift game, for example, is similar to the prisoner's dilemma except that the payoff ranking is $a_{BA}>a_{AA}>a_{AB}>a_{BB}$ instead of $a_{BA}>a_{AA}>a_{BB}>a_{AB}$. As a result, $B$ is no longer a dominant strategy. There are two primary variants of the snowdrift game, one with $\left(a_{AB}+a_{BA}\right) /2<a_{AA}$ and one with $\left(a_{AB}+a_{BA}\right) /2>a_{AA}$. In both cases, collective learning via random encounters within a sub-population results in optimal payoffs. (In the latter case, this is also true for learning within stable pairs.) We also observe optimal outcomes in stag hunt games ($a_{AA}>a_{BA}>a_{BB}>a_{AB}$) provided $\left(a_{AB}+a_{BA}\right) /2>a_{BB}$. See \fig{GP} for details.

\begin{figure}
\centering
\includegraphics[width=0.9\textwidth]{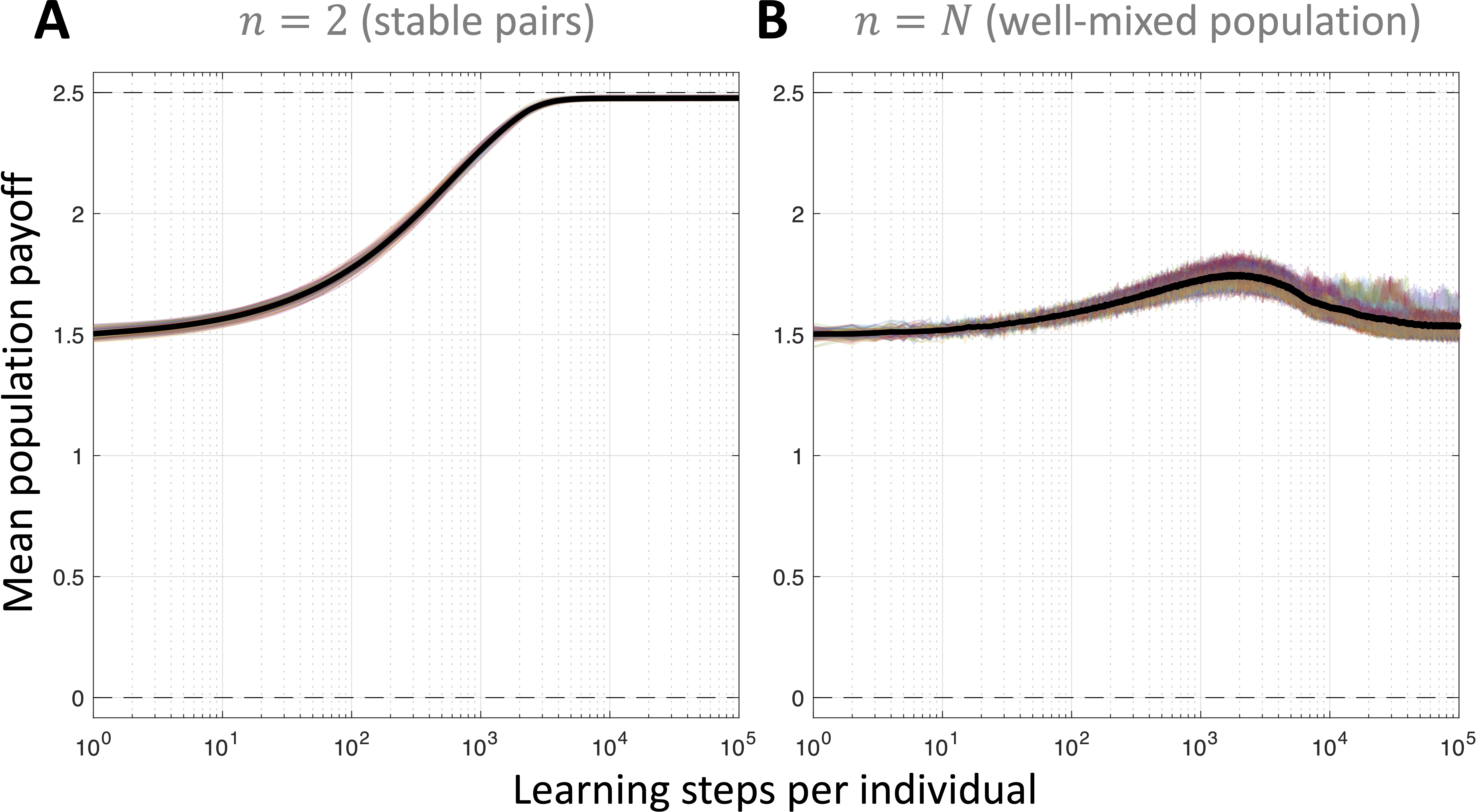}
\caption{\textbf{Pairwise versus collective learning in the battle of the sexes game.} In contrast to repeated prisoner's dilemmas, learning in stable pairs leads to near-optimal outcomes for a population in the repeated battle of the sexes game, whereas collective learning has difficulty achieving and maintaining a high mean payoff. In both panels, the population size is $N=1{,}000$, the number of runs is $100$, the game parameters are $\left(a_{AA},a_{AB},a_{BA},a_{BB}\right) = \left(1,3,2,0\right)$, the continuation probability is $\lambda =0.9999$, and the learning rate is $\eta =10^{-3}$.\label{fig:BSG}}
\end{figure}

A population of selfish, ephemeral learners is not optimal for all types of games, however. The battle of the sexes game \citep{maynardsmith:CUP:1982,luce:D:1989} provides a contrast to the preceding results. This game is characterized by two players with competing preferences for what to do (e.g.~what kind of event to attend). We consider a symmetric version of this game (also known as the ``hero'' game \citep{tanimoto:S:2015}), in which the action $A$ may be thought of as ``go with own preference'' while $B$ is ``go with partner's preference.'' But they would prefer to be together than apart, which leads to a coordination game with payoff ranking $a_{AB}>a_{BA}>a_{AA}>a_{BB}$. \fig{BSG}\textbf{A} shows how learning in stable pairs quickly brings a population close to its optimal average payoff. But this is not the case for a population of randomly-interacting learners (\fig{BSG}\textbf{B}), which may be attributed to the difficulty of achieving consistent anti-coordination when learning partners are not fixed. We note that this kind of anti-coordination is different from that of the alternating prisoner's dilemma (\fig{alternatingPD}). Due to the nature of the underlying one-shot game, a pair of selfish learners in the battle of the sexes games tends to arrive at asymmetric outcomes, with one player receiving $a_{AB}$ and the other getting $a_{BA}$. In contrast, the policies of anti-coordination in the alternating prisoner's dilemma gives both players $\left(a_{AB}+a_{BA}\right) /2$. The average pair payoff is $\left(a_{AB}+a_{BA}\right) /2$ in both cases, but in the former it is more difficult to achieve a policy of anti-coordination in a population.

Finally, there are both prisoner's dilemma and stag hunt games satisfying $\left(a_{AB}+a_{BA}\right) <a_{BB}$. For these games, neither learning in stable pairs nor collective learning efficiently achieves a maximum average payoff (see \fig{LP}). As a result, across a variety of different kinds of games, learning in stable pairs is not universally better than collective learning or vice versa. But they do generally result in very different outcomes -- and collectively learning can often produce optimal outcomes where learning in stable pairs does not.

\subsection{Eulerian perspective: tracking distributions of agents}
Our goal has been to compare two extremes: learning in stable pairs and collective learning in large populations. So far, we have focused on large but finite populations and an infinite strategy space. We now turn to a dual approach, using an infinite population and a discretized, finite strategy space. Specifically, we focus on a single sub-population ($n=N\rightarrow\infty$), motivated by a desire to understand the extreme case in which two learners never interact more than once.

Let $\rho$ be the density of memory-one strategies in the infinite population. Although the space of memory-one strategies may be identified with $\left[0,1\right]^{5}$, it is convenient to think of $\rho$ as a density on $\mathbb{R}^{5}$ that is supported on $\left[0,1\right]^{5}$. At time $t$, the gradient direction of an individual using strategy $\p$ is a random variable, $\bm{V}\left[\rho\right]\left(\p\right)$, with $\rho\left(\q ,t\right)$ the density of direction $\nabla_{\p}\pi\left(\p ,\q\right)$. This gradient field requires defining the payoff, $\pi$, outside of $\left[0,1\right]^{5}$, but the extension of $\pi$ to $\mathbb{R}^{5}$ is irrelevant due to the support of $\rho$.

To ensure that the support of the density is always contained in $\left[0,1\right]^{5}$ for all $t$ whenever it is at time $t=0$, we need a modified gradient field of the form $\widetilde{\bm{V}}=W\bm{V}$ for some matrix $W$, which represents the projection operator in \eq{PGA}. We wish for $W$ to be the identity matrix away from the boundary, but for it to ensure the gradient is always inward pointing on $\left[0,1\right]^{5}$ to retain the salient features of the search process. We take $W$ to be a diagonal matrix depending on both $\p$ and $\q$ (the partner strategy of $\p$) on the boundary, with $W_{ii}=1$ unless either \emph{(i)} $p_{i}=0$ and $\left(\nabla_{\p}\pi\left(\p ,\q\right)\right)_{i}<0$ or \emph{(ii)} $p_{i}=1$ and $\left(\nabla_{\p}\pi\left(\p ,\q\right)\right)_{i}>0$, in which case $W_{ii}=0$.

With this modified gradient field, we consider a continuous-time process in which a learner randomly chooses a partner, computes the gradient direction $\widetilde{\bm{V}}\left[\rho\right]\left(\p^{t}\right)$, and then lets its strategy at time $t+\Delta t$ be $\p^{t+\Delta t}=\p^{t}+\left(\Delta t\right)\widetilde{\bm{V}}\left[\rho\right]\left(\p^{t}\right)$. Thus, a time step of $\Delta t=\eta$ corresponds to the discrete-time model presented previously. In the limit as $\Delta t\rightarrow 0$, the strategy density evolves according to the advection equation
\begin{align}
\frac{\partial\rho\left(\p ,t\right)}{\partial t} &= - \nabla_{\p} \cdot \left( \rho\left(\p ,t\right) \int_{\mathbf{q}\in\mathbb{R}^{5}} W\left(\p ,\q\right)\nabla_{\mathbf{p}}\pi\left(\mathbf{p},\mathbf{q}\right) \rho\left(\mathbf{q},t\right) \, d\mathbf{q} \right) . \label{eq:PDE}
\end{align}
Notably, this partial differential equation depends on only the mean (projected) gradient direction, which is important because in the mechanistic description of the model we do not require an individual to interact with a large subset of the population. There is no notion of a ``mean'' player in the definition of the model, and only one encounter takes place per individual prior to each learning step. We include a more detailed description of the Eulerian approach in \sect{sm_pde}. In addition to the PDE, we also derive the large-$N$ limit of the system describing the dynamics in discrete time (\eq{mui_deltat}).

The trade-off of this approach is that, to work with this equation numerically, we must discretize the strategy space. To simplify this numerical problem, we study lower-dimensional strategy spaces. If $\p$ is a ``reactive'' strategy, then $p_{xy}$ depends on $y$ only and not on $x$ \citep{kalai:IJGT:1988}. Thus, $p_{AA}=p_{BA}\eqqcolon p_{A}$ and $p_{AB}=p_{BB}\eqqcolon p_{B}$, so $\p$ is specified by a triplet, $\left(p_{0},\left(p_{A},p_{B}\right)\right)\in\left[0,1\right]^{3}$. If we also assume that the game length is infinite (taking the limit $\lambda\rightarrow 1$), then the initial probability of playing $A$ is irrelevant and payoffs depend on $\p =\left(p_{A},p_{B}\right)\in\left[0,1\right]^{2}$ (see \eq{reactive_payoffs}).

Reducing the sophistication of strategies for repeated games also limits both transient dynamics and long-term outcomes. For example, the rigidity of reactive strategies makes it more difficult to efficiently achieve policies of alternating cooperation in prisoner's dilemma games with $a_{AA}<\left(a_{AB}+a_{BA}\right) /2$ (c.f. \fig{alternatingPD}). But reactive strategies do preserve the qualitative behavior seen in the standard prisoner's dilemma of \fig{standardPD}, and so we can use this simpler strategy space to study collective learning from a Eulerian perspective.

\begin{figure}
\centering
\includegraphics[width=0.5\textwidth]{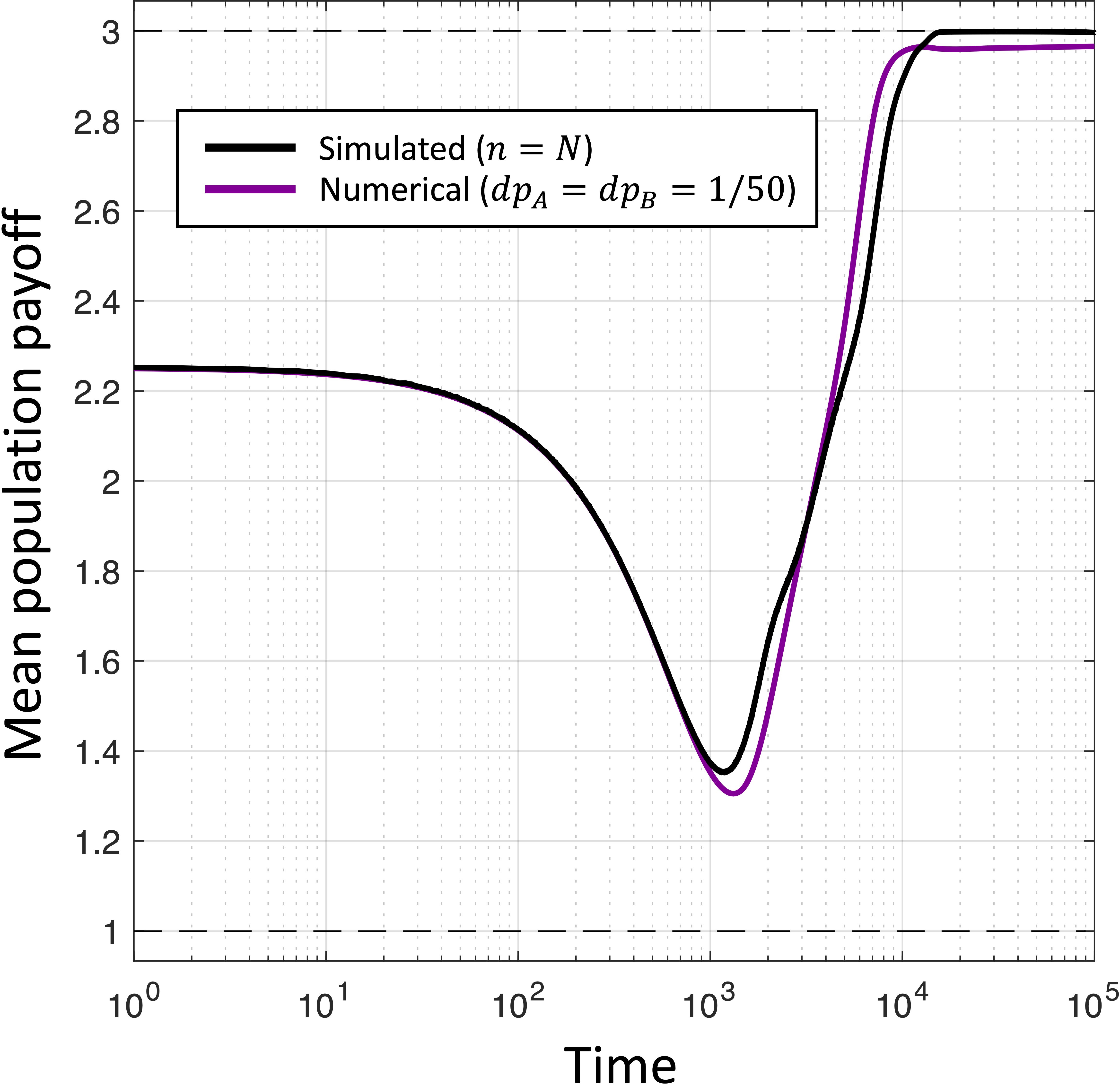}
\caption{\textbf{Learning dynamics of reactive strategies in an infinite population.} Solving \eq{PDE} numerically (\fig{numerical_scheme}), we plot the mean population payoff over time when players use a simple class of reactive strategies. The mean payoff (purple) predicted by the PDE closely tracks the mean payoff observed in agent-based simulations (black) when $n=N=1{,}000$. The payoffs in the numerical scheme end up slightly below the maximum payoff of $3$, due to the use of a finite grid for the strategy space, obtained by dividing $\left[0,1\right]$, the domain of each coordinate $p_{A}$ and $p_{B}$, into $50$ equally spaced subintervals. The time scale is chosen such that one unit of time in the continuous model corresponds to one time step in the discrete model.\label{fig:PDE}}
\end{figure}

\fig{numerical_scheme} illustrates the numerical scheme to compute how $\rho$, the density of strategies, changes over time. In \fig{PDE}, we see results from the PDE that are in close qualitative agreement with individual-based (that is, Lagrangian) simulations. In addition to providing a different perspective on the learning problem (involving a different kind of calculation), the Eulerian approach also illustrates that the learning timescale does not diverge as $N\rightarrow\infty$ in repeated prisoner's dilemmas. In the Lagrangian (individual-based) approach, we observed relatively quick convergence to the mean payoff maximum in prisoner's dilemma games when $n=N\gg 0$, and indeed this is also reflected in the solution of the Eulerian PDE (\fig{PDE}).

\begin{figure}
\centering
\includegraphics[width=\textwidth]{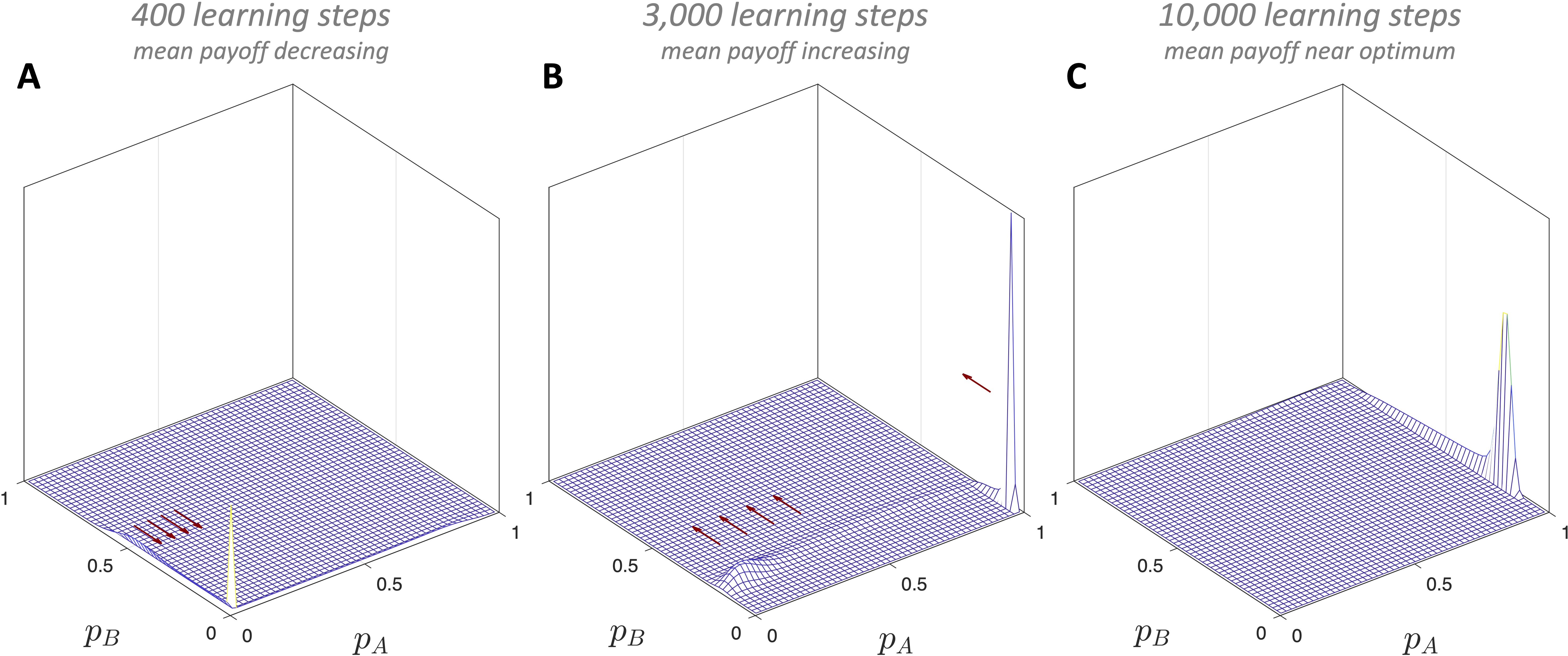}
\caption{\textbf{Snapshots of the strategy density.} Panels \textbf{A}, \textbf{B}, and \textbf{C} depict the strategy density in \vid{density} after $400$, $3{,}000$, and $10{,}000$ learning steps, respectively. Following an initial attraction to defection (\textbf{A}), individuals near $\p =\left(1,0\right)$ who reciprocate cooperation and punish defection can incentivize their interaction partners to cooperate. As a result, we observe an outflux from defection in \textbf{B}, which increases the mean population payoff. Finally, as the mean population payoff settles near its optimal value ($3$, in this example), strategies tend to reciprocate cooperation but forgive some amount of defection ($p_{B}$ positive but not too large). Note that strategies develop a limited amount of forgiveness in order to avoid being exploited.\label{fig:density_snapshots}}
\end{figure}

The evolution of the density leading to \fig{PDE} is depicted in \vid{density}, as well as in snapshots in \fig{density_snapshots}. This density reveals that the initial dip in mean payoff is due to a general trend toward unconditional defection ($\p =\left(0,0\right)$); see \fig{density_snapshots}\textbf{A}. The resulting decline in payoffs is consistent with what is seen for pairs (\fig{standardPD}\textbf{A} and \fig{standardPD_pairs}) following a random initial state. However, some of the density remains concentrated near $\left(1,0\right)$. Since these strategies reciprocate cooperation and punish defection, they are capable of incentivizing cooperation in selfish agents, even in those who tend to defect with high probability. As a result, we observe a wave develop outwards from unconditional defection toward more cooperative strategies (\fig{density_snapshots}\textbf{B}), which increases the mean population payoff. And even though some of these incentivizing strategies are present in the early stages of the process, at that time most individuals are likely to interact with strategies that cannot elicit cooperation; as a result, learners, on average, tend to decrease their cooperation levels more frequently than they increase them, leading to a net degradation of payoff. Once defection is predominant, individuals near $\left(1,0\right)$ benefit from being both more cooperative (increasing $p_{A}$) and more forgiving (increasing $p_{B}$). This, in turn, incentivizes cooperation in the rest of the population, provided there is not too much forgiveness for defection (\fig{density_snapshots}\textbf{C}).

\section{Discussion}
In this study, we have compared selfish learning with stable pairs to selfish learning with stochastic encounters in a population to determine \emph{(i)} whether there are differences in outcomes between these two learning processes and \emph{(ii)} whether observed differences are robust across all classes of interactions. The result of our investigation can be broadly summarized as saying ``yes'' to question \emph{(i)} and ``no'' to question \emph{(ii)}.

Our study is related to several existing lines of inquiry in the literature on learning and evolutionary dynamics. The first is evolutionary dynamics in a single sub-population ($n=N$), where similar strategic trajectories have been noticed in populations of replicators who ``learn" by copying more successful individuals. In that setting, \citet{nowak:Nature:1992} observed a transition from random reactive strategies, to unconditional defection, and then to tit-for-tat (copy the opponent), and finally to generous tit-for-tat (copy the opponent, but forgive errors). Indeed, the population payoff trajectories found in that context \citep[][Fig.~1]{nowak:Nature:1992} resemble the qualitative behavior seen in \fig{standardPD} and \fig{PDE}. However, in stark contrast to the replicators of \citet{nowak:Nature:1992}, our model is based on individual learning through steps of gradient ascent (locally rational, selfish updates), as opposed biased imitation of fitter individuals. In our setting, when two players meet, neither one replicates the strategy of the other, regardless of how successful the two strategies are; in fact, the learning process may bring a player's strategy further away from their opponent's. This is a fundamental difference between the learning process we consider here and models based on biased replicators. Although biased replication of successful strategies can be seen as a simplistic form of learning, the process that we study is completely different in that it requires individuals to have cognitive ability, to be aware of the payoffs in the game they are playing, and to compute how they can improve their payoff against their current opponent by a small change in strategy.

There is another important distinction between our study and the literature on traditional replicator models of strategy dynamics: whereas populations are frequently used to study the evolution of cooperation \citep{axelrod:APSR:1981,nowak:Nature:2004,santos:PRL:2005,nowak:Science:2006,szabo:PR:2007,moyano:JTB:2009}, our focus is decidedly not on cooperation per se. Cooperation is an action in several of the games we consider, but we are concerned with achieving optimal payoffs for a population, regardless of whether those are achieved by cooperation or not. Indeed, this is one reason for studying the version of the prisoner's dilemma shown in \fig{alternatingPD}, where mutual cooperation is actually suboptimal and collective learning nonetheless achieves the optimal outcome.

Furthermore, we note the distinction between the deterministic formulation described in \eq{PDE} and evolutionary invasion analysis \cite{geritz:EE:1998}, which also uses gradient-based dynamics. The latter, sometimes referred to as ``adaptive dynamics,'' concerns the invasion of initially-rare mutant traits in a resident population. Our approach, by contrast, allows for multiple types coexisting in the population, and so it is more akin to the model of population games of \citet{friedman:JET:2013} than to invasion analysis, except we use a higher-dimensional strategy space due to the complexity of repeated games. \vid{density} and \fig{density_snapshots} demonstrate how the mechanics of strategy evolution in our learning model differ significantly from those of both evolutionary \citep[][Fig. 1]{nowak:Nature:1992} and adaptive \citep[][Fig. 6]{nowak:Science:2004} dynamics.

Our study is also related to the literature on multi-agent learning. For many problems, the goal might be to understand the effects of learning rules when encounters are decidedly stochastic and ephemeral. Player $i$ might never meet player $j$ again, but the next partner will have faced someone \emph{like} player $i$ in the past. An agent learns from past experiences and adjusts, bringing new behavior into future encounters. In our example of the repeated prisoner's dilemma, we have seen that moving from pairwise stable learning to population-based collective learning allows selfish learning to be quite efficient at attaining optimal outcomes for all. But these findings are somewhat sensitive to the interaction type (c.f. \fig{BSG}) as well as to the exact implementation of the learning rules themselves, and so we make no claim that collective learning is universally more efficient than learning in stable pairs. 

Instead, we emphasize that the population-based model is highly relevant to many real-world situations of multi-agent learning, and the outcomes of this learning process depend dramatically on how likely one is to re-encounter the same learning partner. Of course, models of multi-agent learning can be more complex, involving dynamic environments \cite{akiyama:PD:2000,sato:PD:2005,barfuss:PRE:2019}. Even social dilemmas can be extended spatially and temporally \citep{leibo:AAMAS:2017}. Nevertheless, the simplicity of our model, based on repeated matrix games with purely myopic, selfish strategy updates, has some advantages. To understand multi-agent learning in more realistic application cases, we must also understand the effects of a population in simpler (even if idealized) settings. The complexity introduced by stochastic encounters in a population makes this problem difficult even for repeated matrix games, but here it is still tractable. An important area of future research will be to study this problem in more complicated settings, including spatial and temporal extensions.

From our study of repeated matrix games, it is clear that the outcomes of learning in a population will depend on the initial distribution to a significant degree. If all individuals were to start with deterministic policies of playing $B$ (defecting) in every round of a prisoner's dilemma, then no amount of flexibility in partner choice could bring the population out of this state under selfish learning. The distributions we consider (arcsine and uniform) both introduce a large diversity of strategies in the initial population prior to any learning, which we view as important determinants of long-term prosperity. However, we note that even for stable pairs in the iterated prisoner's dilemma, the long-run outcomes cannot be predicted from the initial level of cooperation alone. For example, if two players both initially use unconditional cooperation (``ALLC''), then their optimization process will lead them to mutual defection; but if they both use the strategy tit-for-tat (``TFT''), then they remain at mutual cooperation. This property holds even though ALLC against ALLC results in the same initial level of cooperation as TFT against TFT.

We have focused on repeated games that are sufficiently long because these games are known to have rich spaces of equilibria \citep{fudenberg:E:1986}. While our concern has been payoffs (in particular, mean population payoffs) rather than Nash or subgame-perfect equilibria, long time horizons ensure that current behaviors can be punished or rewarded in the future with high probability, and thus, in principle, there are interesting strategies available for players to learn in the first place. In many classes of interactions, interesting strategies exist even when the length of the game is slightly shorter, and an open question is how the length of the game influences the learning process.

Finally, the population size ties into both the initial distribution and the length of interactions, and it may have effects on long-term learning outcomes. When individuals choose initial strategies at random, larger population sizes lead to a broader variety of behaviors present prior to learning. The effect of this diversity is almost trivial if individuals interact in stable pairs since the mean population payoff then just converges to the expected payoff for a single pair as the population grows (c.f.~\fig{standardPD}\textbf{A} and \fig{standardPD_pairs}). But with larger sub-populations, especially when any two individuals in a population can interact, this strategy variety matters a lot. In such situations, individuals are also less likely to interact in two subsequent time steps if $N$ is large, which leaves open the possibility that somewhat large mean payoffs are possible even for shorter games.

\begin{center}
\Large\textbf{Appendix}
\end{center}

\setcounter{equation}{0}
\setcounter{figure}{0}
\setcounter{section}{0}
\setcounter{table}{0}
\renewcommand{\thesection}{A.\arabic{section}}
\renewcommand{\thesubsection}{A.\arabic{section}.\arabic{subsection}}
\renewcommand{\theequation}{A.\arabic{equation}}
\renewcommand{\thetable}{A.\arabic{table}}
\renewcommand{\figurename}{\footnotesize Supplementary~Figure}
\renewcommand{\tablename}{\footnotesize Supplementary~Table}

\section{Repeated games and payoffs}\label{sec:sm_payoffs}
For $\p ,\q\in\left[0,1\right]^{5}$, $\mathbf{u}\in\mathbb{R}^{4}$, and $\lambda\in\left[0,1\right)$, let $M\left(\p ,\q ,\mathbf{u}\right)$ denote the matrix
\begin{align}
\begin{pmatrix}
\left(1-\lambda\right) p_{0}q_{0}+\lambda p_{AA}q_{AA}-1 & \left(1-\lambda\right) p_{0}+\lambda p_{AA}-1 & \left(1-\lambda\right) q_{0}+\lambda q_{AA}-1 & u_{1} \\
\left(1-\lambda\right) p_{0}q_{0}+\lambda p_{AB}q_{BA} & \left(1-\lambda\right) p_{0}+\lambda p_{AB}-1 & \left(1-\lambda\right) q_{0}+\lambda q_{BA} & u_{2} \\
\left(1-\lambda\right) p_{0}q_{0}+\lambda p_{BA}q_{AB} & \left(1-\lambda\right) p_{0}+\lambda p_{BA} & \left(1-\lambda\right) q_{0}+\lambda q_{AB}-1 & u_{3} \\
\left(1-\lambda\right) p_{0}q_{0}+\lambda p_{BB}q_{BB} & \left(1-\lambda\right) p_{0}+\lambda p_{BB} & \left(1-\lambda\right) q_{0}+\lambda q_{BB} & u_{4} \\
\end{pmatrix} .
\end{align}
With $\mathbf{u}_{X}\coloneqq\left(a_{AA},a_{AB},a_{BA},a_{BB}\right)$ and $\mathbf{u}_{Y}\coloneqq\left(a_{AA},a_{BA},a_{AB},a_{BB}\right)$, the expected payoffs to the two players in the repeated game can then be written explicitly as the following quotients \citep{press:PNAS:2012,mamiya:PRE:2020}:
\begin{subequations}
\begin{align}
\pi_{X}\left(\p ,\q\right) &= \frac{\det M\left(\p ,\q ,\mathbf{u}_{X}\right)}{\det M\left(\p ,\q ,\mathbf{1}\right)} ; \\
\pi_{Y}\left(\p ,\q\right) &= \frac{\det M\left(\p ,\q ,\mathbf{u}_{Y}\right)}{\det M\left(\p ,\q ,\mathbf{1}\right)} .
\end{align}
\end{subequations}
Since $\pi_{Y}\left(\p ,\q\right) =\pi_{X}\left(\q ,\p\right)$ due to the symmetry of the game, all payoffs can be expressed in terms of $\pi_{X}$, which for simplicity we denote by just $\pi$. As long as $\lambda\in\left[0,1\right)$ is fixed, this function is well-defined and bounded on all of $\left[0,1\right]^{5}$, as are its partial derivatives.

In the numerical scheme depicted in \fig{numerical_scheme}, we use reactive strategies to reduce the dimension of the strategy space. If $\p =\left(p_{A},p_{B}\right)$ and $\q =\left(q_{A},q_{B}\right)$ are reactive strategies for $X$ and $Y$, then, in the limit $\lambda\rightarrow 1$,
\begin{align}
\pi\left(\p ,\q\right) &= \frac{\det
\begin{pmatrix}
p_{A}q_{A}-1 & p_{A}-1 & q_{A}-1 & a_{AA} \\
p_{A}q_{B} & p_{A}-1 & q_{B} & a_{AB} \\
p_{B}q_{A} & p_{B} & q_{A}-1 & a_{BA} \\
p_{B}q_{B} & p_{B} & q_{B} & a_{BB} \\
\end{pmatrix}}{\det
\begin{pmatrix}
p_{A}q_{A}-1 & p_{A}-1 & q_{A}-1 & 1 \\
p_{A}q_{B} & p_{A}-1 & q_{B} & 1 \\
p_{B}q_{A} & p_{B} & q_{A}-1 & 1 \\
p_{B}q_{B} & p_{B} & q_{B} & 1 \\
\end{pmatrix}} \nonumber \\
&= \frac{\left(\substack{\left(p_{A}q_{B}+p_{B}\left(1-q_{B}\right)\right)\left(p_{B}q_{A}+\left(1-p_{B}\right) q_{B}\right) a_{AA} \\ + \left(p_{A}q_{B}+p_{B}\left(1-q_{B}\right)\right)\left(1-p_{A}q_{A}-\left(1-p_{A}\right) q_{B}\right) a_{AB} \\ +\left(p_{B}q_{A}+\left(1-p_{B}\right) q_{B}\right)\left(1-p_{A}q_{A}-p_{B}\left(1-q_{A}\right)\right) a_{BA} \\ +\left(1-p_{A}q_{A}-p_{B}\left(1-q_{A}\right)\right)\left(1-p_{A}q_{A}-\left(1-p_{A}\right) q_{B}\right) a_{BB}}\right)}{\left(1-\left(p_{A}-p_{B}\right)\left(q_{A}-q_{B}\right)\right)^{2}} . \label{eq:reactive_payoffs}
\end{align}
The only time in which this payoff is not well-defined is when $\p =\q =\left(1,0\right)$ or $\p =\q =\left(0,1\right)$. In these cases, the Markov chain defining the state transitions has more than one stationary distribution, so an initial action is also needed to determine final payoffs. However, this does not impact the numerical scheme shown in \fig{numerical_scheme} (or the results shown in \fig{PDE}) because neither payoffs nor gradient directions are evaluated at $\p =\q =\left(1,0\right)$ or $\p =\q =\left(0,1\right)$.

\section{Strategy densities in the large-population limit}\label{sec:sm_pde}
Here, we give a sketch of the derivation of \eq{PDE} in the main text. We consider the general case in which strategies have $m$ different components, thought of as elements of $\left[0,1\right]^{m}\subseteq\mathbb{R}^{m}$. To simplify notation slightly in this setting, we let $S=\mathbb{R}^{m}$ and we let $s_{i}\in S$ denote the strategy of individual $i$ (in place of the standard notation, $\p$ and $\q$, which we used previously when $m=5$).

For a finite population of size $N$, consider the set of partnership maps
\begin{align}
\Gamma_{N} &\coloneqq \left\{ \gamma : \left\{1,\dots ,N\right\}\rightarrow\left\{1,\dots ,N\right\} \ \mid\ \gamma\left(i\right)\neq i,\ \left(\gamma\circ\gamma\right)\left(i\right) =i\textrm{ for }i=1,\dots ,N\right\} .
\end{align}
Informally, the value $\gamma\left(i\right)$ is the partner of $i$ chosen in one time step. The condition $\gamma\left(i\right)\neq i$ ensures that my partner is someone other than myself, and the condition $\left(\gamma\circ\gamma\right)\left(i\right) =i$ ensures that the partner of my partner is myself. We consider a uniform distribution over all such maps, with each $\gamma\in\Gamma_{N}$ being chosen with probability $1/\left|\Gamma_{N}\right|$. Sampling at one time step is independent from that of another time step.

Suppose that $\mu^{N}\left(\-- ,t\right)$ is the distribution on $S^{N}$ at time $t\in\left\{0,1,2,\dots\right\}$. Note that the support of this distribution is contained in $\left[0,1\right]^{m}\subseteq S$. Let $\mathcal{F}\left(S^{N}\right)$ denote the Borel $\sigma$-algebra on $S^{N}$, which contains all of the open and closed subsets of $S^{N}$ and is used to speak of $\mu^{N}$ as a probability measure on $S^{N}$. For $s\in S^{N}$, $E\in\mathcal{F}\left(S^{N}\right)$, and $\gamma\in\Gamma_{N}$, let $\kappa^{\gamma}\left(s,E\right)$ denote whether the state is in $E$ in the next time step, given that the current state is $s$ and the partnership map is $\gamma$. A one-step analysis of the Markov chain gives
\begin{align}
\mu^{N}\left(E,t+1\right) &= \sum_{\gamma\in\Gamma_{N}} \frac{1}{\left|\Gamma_{N}\right|} \int_{s\in S^{N}} \kappa^{\gamma}\left(s,E\right) \, \mu^{N}\left(ds,t\right) .
\end{align}
If $E=E_{1}\times\cdots\times E_{N}$ is a measurable rectangle, where $E_{i}\in\mathcal{F}\left(S\right)$ for $i=1,\dots ,N$, then, by the definition of the learning process, we can decompose $\kappa^{\gamma}\left(s,E\right)$ into a product,
\begin{align}
\kappa^{\gamma}\left(s,E\right) &= \prod_{i=1}^{m} \kappa_{s_{\gamma\left(i\right)}}\left(s_{i},E_{i}\right) ,
\end{align}
where
\begin{align}
\kappa_{s_{j}}\left(s_{i},E_{i}\right) &\coloneqq \delta_{\textrm{proj}_{S}\left(s_{i}+\nabla_{s_{i}}\pi\left(s_{i},s_{j}\right)\right)}\left(E_{i}\right) .
\end{align}

For $I\subseteq\left\{1,\dots ,N\right\}$, consider now the marginal distribution induced by the projection $S^{N}\rightarrow S^{I}$, which we denote by $\mu_{I}^{N}$. This distribution tracks the strategies in the set $I$, ignoring all others in the population. On measurable rectangles,
\begin{align}
\mu_{I}^{N}\left(\prod_{i\in I}E_{i},t+1\right) &= \sum_{\gamma\in\Gamma_{N}} \frac{1}{\left|\Gamma_{N}\right|} \int_{s\in S^{N}} \kappa^{\gamma}\left(s,\prod_{i\in I}E_{i}\right) \, \mu^{N}\left(ds,t\right) \nonumber \\
&= \sum_{\gamma\in\Gamma_{N}} \frac{1}{\left|\Gamma_{N}\right|} \int_{s\in S^{N}} \prod_{i\in I} \kappa_{s_{\gamma\left(i\right)}}\left(s_{i},E_{i}\right) \, \mu^{N}\left(ds,t\right) \nonumber \\
&= \sum_{\gamma\in\Gamma_{N}} \frac{1}{\left|\Gamma_{N}\right|} \int_{s\in S^{I\cup\gamma\left(I\right)}} \prod_{i\in I} \kappa_{s_{\gamma\left(i\right)}}\left(s_{i},E_{i}\right) \, \mu_{I\cup\gamma\left(I\right)}^{N}\left(ds,t\right) . \label{eq:mu_marginal}
\end{align}
Since the sequence of strategies in the population is exchangeable, we have $\mu_{I}^{N}=\mu_{\tau\left(I\right)}^{N}$ for any bijection $\tau :\left\{1,\dots ,N\right\}\rightarrow\left\{1,\dots ,N\right\}$. Thus, there is a unique measure $\widetilde{\mu}_{k}^{N}$ for each subset size, $k=\left| I\right|$. If $q_{k,\ell}^{N}$ denotes the probability, when $\left| I\right| =k$, of choosing $\gamma\in\Gamma_{N}$ such that $\left| I\cup\gamma\left(I\right)\right| =\ell$, then \eq{mu_marginal} reduces to
\begin{align}
\widetilde{\mu}_{k}^{N} \left(\prod_{i=1}^{k}E_{i},t+1\right) 
&= \sum_{\ell =k}^{2k} q_{k,\ell}^{N} \int_{s\in S^{\ell}} \left[\prod_{i=1}^{\ell -k} \kappa_{s_{k+i}}\left(s_{i},E_{i}\right) \prod_{i=\ell -k+1}^{k} \kappa_{k-i+1}\left(s_{i},E_{i}\right)\right] \, \widetilde{\mu}_{\ell}^{N}\left(ds,t\right) . \label{eq:mu_tilde}
\end{align}

Our first goal is to understand what the discrete-time process looks like in the $N\rightarrow\infty$ limit. Specifically, we would like to understand the mean population payoff,
\begin{align}
\overline{\Pi}_{t}^{N} &\coloneqq \sum_{\gamma\in\Gamma_{N}} \frac{1}{\left|\Gamma_{N}\right|} \left[ \frac{1}{N} \sum_{i=1}^{N} \int_{\left(s_{i},s_{\gamma\left(i\right)}\right)\in S^{2}} \pi\left(s_{i},s_{\gamma\left(i\right)}\right) \, \mu_{\left\{i,\gamma\left(i\right)\right\}}^{N}\left(d\left(s_{i},s_{\gamma\left(i\right)}\right) ,t\right) \right] .
\end{align}
By the previous observation about symmetry, the mean population payoff can be written as
\begin{align}
\overline{\Pi}_{t}^{N} &= \int_{\left(s_{1},s_{2}\right)\in S^{2}} \pi\left(s_{1},s_{2}\right) \, \mu_{\left\{1,2\right\}}^{N}\left(d\left(s_{1},s_{2}\right) ,t\right) .
\end{align}
So, to understand the $N\rightarrow\infty$ limit of the mean population payoff, $\overline{\Pi}_{t}^{\infty}\coloneqq\lim_{N\rightarrow\infty}\overline{\Pi}_{t}^{N}$, it suffices to understand the distribution $\mu_{\left\{1,2\right\}}^{N}$ as $N\rightarrow\infty$.

We first need to see that the limit $\mu_{\left\{1,2\right\}}^{N}$ as $N\rightarrow\infty$ exists. On the right-hand side of \eq{mu_tilde}, only $q_{k,\ell}$ and $\widetilde{\mu}_{\ell}^{N}$ depend on $N$. Moreover, for any fixed $k$ and $\ell$ with $\ell\geqslant k$, we have $\lim_{N\rightarrow\infty}q_{k,\ell}^{N}=\delta_{\ell ,2k}$. Using \eq{mu_tilde}, one can argue by induction that for any $k$ and $t$, $\lim_{N\rightarrow\infty}\widetilde{\mu}_{k}^{N}\left(\--,t\right)$ exists. (Since the initial distribution for each player is chosen independently from some $\mu_{0}$, we have $\widetilde{\mu}_{\ell}^{N}\left(\--,0\right) =\prod_{i=1}^{\ell}\mu_{0}$, which is independent of $N$ and thus has a limit as $N\rightarrow\infty$. It follows from \eq{mu_tilde} that $\widetilde{\mu}_{\ell}^{N}\left(\--,1\right)$ has a limit as $N\rightarrow\infty$. Iterating this argument gives the result for larger values of $t$.)

Taking the limit of \eq{mu_tilde} as $N\rightarrow\infty$ then gives
\begin{align}
\widetilde{\mu}_{k}^{\infty} \left(\prod_{i=1}^{k}E_{i},t+1\right) 
&= \int_{s\in S^{2k}} \prod_{i=1}^{k} \kappa_{s_{k+i}}\left(s_{i},E_{i}\right) \, \widetilde{\mu}_{2k}^{\infty}\left(ds,t\right) . \label{eq:mu_tilde_infty}
\end{align}
We are primarily interested in $\widetilde{\mu}_{2}^{\infty}$, or, equivalently, $\mu_{\left\{1,2\right\}}^{\infty}$. For $k=1$ and $k=2$, respectively, \eq{mu_tilde_infty} gives
\begin{subequations}
\begin{align}
\widetilde{\mu}_{1}^{\infty}\left(E_{1},t+1\right) &= \int_{s\in S^{2}} \kappa_{s_{2}}\left(s_{1},E_{1}\right) \, \widetilde{\mu}_{2}^{\infty}\left(ds,t\right) \label{eq:mu1_infty} ; \\
\widetilde{\mu}_{2}^{\infty}\left(E_{1}\times E_{2},t+1\right) &= \int_{s\in S^{4}} \kappa_{s_{3}}\left(s_{1},E_{1}\right) \kappa_{s_{4}}\left(s_{2},E_{2}\right) \, \widetilde{\mu}_{4}^{\infty}\left(ds,t\right) . \label{eq:mu2_infty}
\end{align}
\end{subequations}
At time $t=0$, we know that all joint distributions are product distributions. Thus, $\widetilde{\mu}_{4}^{\infty}\left(\-- ,0\right) =\widetilde{\mu}_{2}^{\infty}\left(\--,0\right)\times\widetilde{\mu}_{2}^{\infty}\left(\--,0\right)$. Therefore,
\begin{align}
\int_{s\in S^{4}} &\kappa_{s_{3}}\left(s_{1},E_{1}\right) \kappa_{s_{4}}\left(s_{2},E_{2}\right) \, \widetilde{\mu}_{4}^{\infty}\left(ds,0\right) \nonumber \\
&= \left(\int_{\left(s_{1},s_{3}\right)\in S^{2}} \kappa_{s_{3}}\left(s_{1},E_{1}\right) \, \widetilde{\mu}_{2}^{\infty}\left(d\left(s_{1},s_{3}\right),0\right)\right) \nonumber \\
&\qquad \times\left(\int_{\left(s_{2},s_{4}\right)\in S^{2}} \kappa_{s_{4}}\left(s_{2},E_{2}\right) \, \widetilde{\mu}_{2}^{\infty}\left(d\left(s_{2},s_{4}\right),0\right)\right) .
\end{align}
The left-hand side is equal to $\widetilde{\mu}_{2}^{\infty}\left(E_{1}\times E_{2},1\right)$ by \eq{mu2_infty} and the right-hand side is equal to $\widetilde{\mu}_{1}^{\infty}\left(E_{1},1\right)\widetilde{\mu}_{1}^{\infty}\left(E_{2},1\right)$ by \eq{mu1_infty}. More generally, to see that $\widetilde{\mu}_{2k}^{\infty}\left(\-- ,t\right)=\widetilde{\mu}_{k}^{\infty}\left(\-- ,t\right)\times\widetilde{\mu}_{k}^{\infty}\left(\-- ,t\right)$ for $t\geqslant 1$, it suffices to know that $\widetilde{\mu}_{4k}^{\infty}\left(\-- ,t-1\right)=\widetilde{\mu}_{2k}^{\infty}\left(\-- ,t-1\right)\times\widetilde{\mu}_{2k}^{\infty}\left(\-- ,t-1\right)$. The argument that we gave for $k=1$ works to establish this claim for any finite $t$, based on the assumption that all initial strategy distributions are chosen independently.

Therefore, using this independence at $N=\infty$ and \eq{mu1_infty} gives
\begin{align}
\mu_{\left\{1\right\}}^{\infty}\left(E_{1},t+1\right) &= \int_{s_{1}\in S} \int_{s_{2}\in S} \kappa_{s_{2}}\left(s_{1},E_{1}\right) \, \mu_{\left\{1\right\}}^{\infty}\left(ds_{2},t\right) \, \mu_{\left\{1\right\}}^{\infty}\left(ds_{1},t\right) .
\end{align}
This is, in fact, our main equation of interest--at least in discrete time. However, since we are dealing with small learning rates, $\eta$, it is natural to ask what kind of model arises in a continuous time limit. We assume that in a time interval of length $\Delta t$, the learning rate associated to the gradient step is exactly $\Delta t$. So, our discrete model in the text with a learning rate of $\eta$ corresponds to a fractional time step of $\Delta t=\eta$. In this case, we have
\begin{align}
\mu_{\left\{1\right\}}^{\infty}\left(E_{1},t+\Delta t\right) &= \int_{s_{1}\in S} \int_{s_{2}\in S} \kappa_{s_{2}}^{\Delta t}\left(s_{1},E_{1}\right) \, \mu_{\left\{1\right\}}^{\infty}\left(ds_{2},t\right) \, \mu_{\left\{1\right\}}^{\infty}\left(ds_{1},t\right) , \label{eq:mui_deltat}
\end{align}
where
\begin{align}
\kappa_{s_{j}}^{\Delta t}\left(s_{i},E_{i}\right) &\coloneqq \delta_{\textrm{proj}_{S}\left(s_{i}+\left(\Delta t\right)\nabla_{s_{i}}\pi\left(s_{i},s_{j}\right)\right)}\left(E_{i}\right) .
\end{align}

If $\varphi$ is a test function, then a density, $\rho$ (thought of as a generalized function), for $\mu_{\left\{1\right\}}^{\infty}$ satisfies
\begin{align}
\rho\left(t\right)\left(\varphi\right) &= \int_{s\in S} \varphi\left(s\right) \, \mu_{\left\{1\right\}}^{\infty}\left(ds,t\right) .
\end{align}
Integrating $\varphi$ with respect to $\mu_{\left\{1\right\}}^{\infty}$ and using \eq{mui_deltat} then gives
\begin{align}
&\frac{\rho\left(t+\Delta t\right) -\rho\left(t\right)}{\Delta t}\left(\varphi\right) \nonumber \\
&\quad = \int_{s_{1}\in S} \int_{s_{2}\in S} \frac{\varphi\left(\textrm{proj}_{S}\left(s_{1}+\left(\Delta t\right)\nabla_{s_{1}}\pi\left(s_{1},s_{2}\right)\right)\right) -\varphi\left(s_{1}\right)}{\Delta t} \, \mu_{\left\{1\right\}}^{\infty}\left(ds_{2},t\right) \, \mu_{\left\{1\right\}}^{\infty}\left(ds_{1},t\right) . \label{eq:discrete_pde}
\end{align}
To evaluate the integrand as $\Delta t\rightarrow 0$, we consider the modifier $w\left(s_{1},s_{2}\right)\in\left\{0,1\right\}^{m}$ with
\begin{align}
w\left(s_{1},s_{2}\right)_{k} &\coloneqq 
\begin{cases}
0 & \left(s_{1}\right)_{k}=0,\ \left(\nabla_{s_{1}}\pi\left(s_{1},s_{2}\right)\right)_{k}<0 , \\
& \\
0 & \left(s_{1}\right)_{k}=1,\ \left(\nabla_{s_{1}}\pi\left(s_{1},s_{2}\right)\right)_{k}>0 , \\
& \\
1 & \textrm{otherwise}.
\end{cases}
\end{align}
for $k=1,\dots ,m$. With $\odot$ denoting the Hadamard product, we then see that
\begin{align}
\lim_{\Delta t\rightarrow 0} &\frac{\varphi\left(\textrm{proj}_{S}\left(s_{1}+\left(\Delta t\right)\nabla_{s_{1}}\pi\left(s_{1},s_{2}\right)\right)\right) -\varphi\left(s_{1}\right)}{\Delta t} \nonumber \\
&= \nabla_{s_{1}}\varphi\left(s_{1}\right)\cdot\left(w\left(s_{1},s_{2}\right)\odot\nabla_{s_{1}}\pi\left(s_{1},s_{2}\right)\right) .
\end{align}
Letting $\Delta t\rightarrow 0$ in \eq{discrete_pde} and using the definition of $\rho$ (in a distributional sense), we obtain
\begin{align}
\frac{\partial\rho\left(t\right)}{\partial t}\left(\varphi\right) &= \int_{s_{1}\in S} \int_{s_{2}\in S} \nabla_{s_{1}}\varphi\left(s_{1}\right)\cdot\left(w\left(s_{1},s_{2}\right)\odot\nabla_{s_{1}}\pi\left(s_{1},s_{2}\right)\right) \, \mu_{\left\{1\right\}}^{\infty}\left(ds_{2},t\right) \, \mu_{\left\{1\right\}}^{\infty}\left(ds_{1},t\right) \nonumber \\
&= \int_{s_{1}\in S} \nabla_{s_{1}} \varphi\left(s_{1}\right)\cdot\left[\int_{s_{2}\in S}\left(w\left(s_{1},s_{2}\right)\odot\nabla_{s_{1}}\pi\left(s_{1},s_{2}\right)\right) \, \mu_{\left\{1\right\}}^{\infty}\left(ds_{2},t\right)\right] \, \mu_{\left\{1\right\}}^{\infty}\left(ds_{1},t\right) \nonumber \\
&= - \nabla_{s_{1}}\cdot\left(\rho\left(t\right)\int_{s_{2}\in S}\left(w\left(s_{1},s_{2}\right)\odot\nabla_{s_{1}}\pi\left(s_{1},s_{2}\right)\right) \, \mu_{\left\{1\right\}}^{\infty}\left(ds_{2},t\right)\right)\left(\varphi\right) \nonumber \\
&= - \nabla_{s_{1}}\cdot\left(\rho\left(t\right)\mathbb{E}_{\rho\left(t\right)}\left[w\left(s_{1},\--\right)\odot\nabla_{s_{1}}\pi\left(s_{1},\--\right)\right]\right)\left(\varphi\right) .
\end{align}
Therefore, by slightly abusing notation and writing $\rho$ as a function, the density evolves according to the advection equation
\begin{align}
\frac{\partial\rho\left(s_{1},t\right)}{\partial t} &= - \nabla_{s_{1}}\cdot\left(\rho\left(s_{1},t\right)\int_{s_{2}\in S}\left(w\left(s_{1},s_{2}\right)\odot\nabla_{s_{1}}\pi\left(s_{1},s_{2}\right)\right) \rho\left(s_{2},t\right) \, ds_{2}\right) .
\end{align}
In particular, from this equation we can obtain the mean population payoff at time $t$,
\begin{align}
\overline{\Pi}_{t}^{\infty} &= \int_{s_{1}\in S} \left(\int_{s_{2}\in S} \pi\left(s_{1},s_{2}\right) \rho\left(s_{2},t\right) \, ds_{2}\right) \rho\left(s_{1},t\right) \, ds_{1} .
\end{align}

\section*{Acknowledgments}
We are grateful to Daniel Cooney, Christian Hilbe, Julian Kates-Harbeck, Eleni Katifori, Andrea Liu, Truong-Son Van, and the soft matter group at the University of Pennsylvania for comments on this manuscript and for helpful conversations related to collective learning. This work was supported by the Simons Foundation (Math+X grant to Y.M.), the National Science Foundation (grant DMS-2042144 to Y.M.), the David \& Lucile Packard Foundation (J.B.P.), and the John Templeton Foundation (J.B.P., grant 62281).

\makeatletter
\@fpsep\textheight
\makeatother

\newpage

\begin{figure}
\centering
\includegraphics[width=0.5\textwidth]{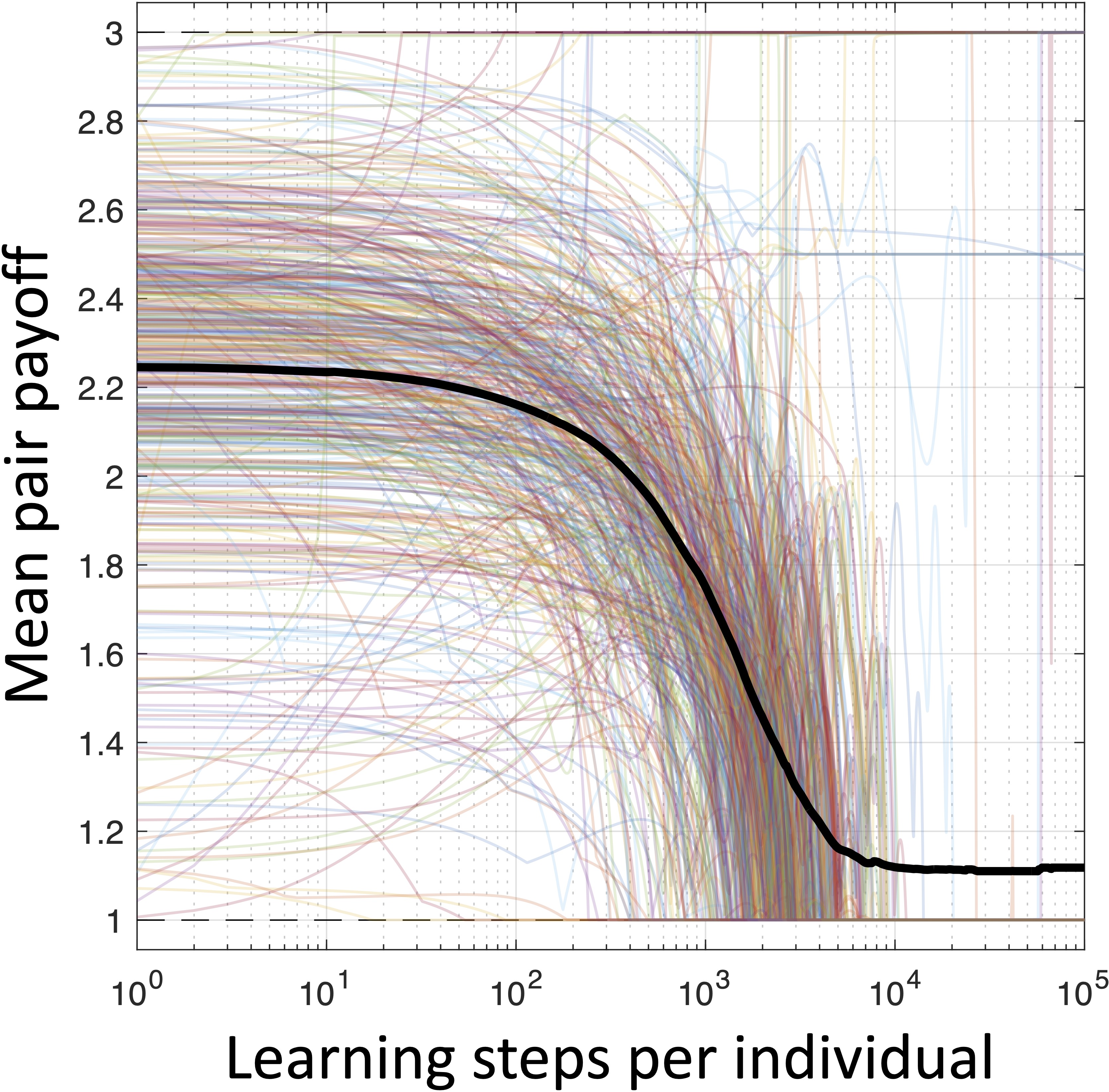}
\caption{\textbf{Mean payoffs of pairs in a prisoner's dilemma.} As an alternative depiction of stable learning in pairs, we can track the mean payoff of each pair rather than the population mean. All parameters are identical to those of \fig{standardPD}\textbf{A}. The mean over $d=500$ trajectories of pairs is shown in black, which corresponds to a single colored trajectory in \fig{standardPD}\textbf{A}. Although this depiction also shows the harmful effects of selfish optimization in stable pairs, we use a different point of comparison in the main text (\fig{standardPD}\textbf{A}) to ensure that the population size is fixed when comparing learning in stable pairs to collective learning.\label{fig:standardPD_pairs}}
\end{figure}

\newpage

\begin{figure}
\centering
\includegraphics[width=0.9\textwidth]{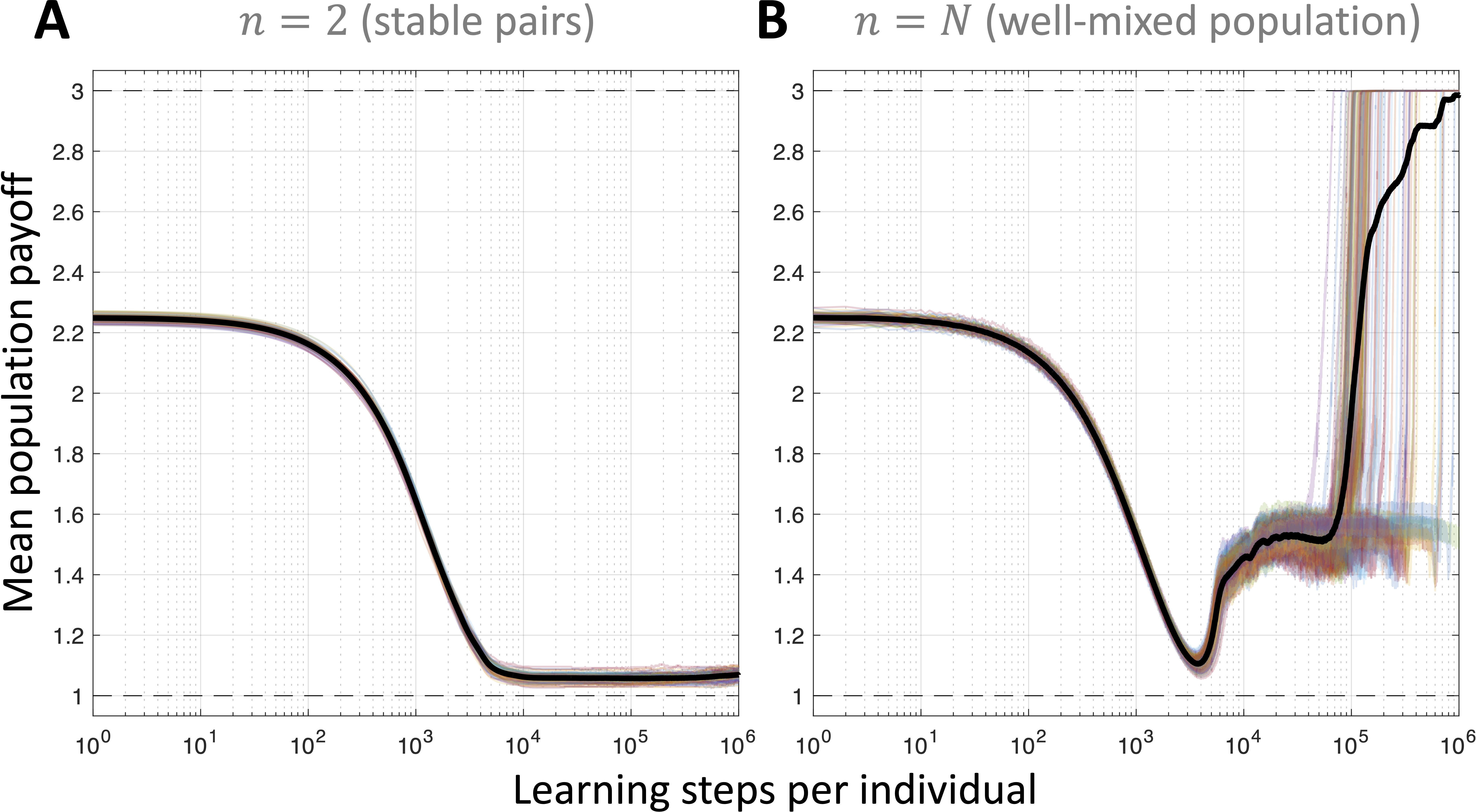}
\caption{\textbf{Pairwise versus collective learning in a standard prisoner's dilemma with uniform initialization.} In the main text, we considered learning rules in which each coordinate of the initial strategy is chosen from an arcsine distribution (to more efficiently explore the corners of $\left[0,1\right]$). Here, we show an analogue of \fig{standardPD} when each coordinate of an initial strategy is chosen from a uniform distribution on $\left[0,1\right]$. The qualitative findings reported in the main text remain the same: stable pairwise learning leads to significantly suboptimal outcomes, while collective learning leads to mean populations close to the maximum possible value. It is worth noting that the trajectories in \textbf{B} are significantly impacted by the choice of initial distribution. In \fig{standardPD}\textbf{B}, we observed payoffs near the maximum value of $3$ after $10^{5}$ learning steps, whereas here it takes approximately $10^{6}$ steps. Furthermore, the payoffs for stable pairwise learning in \textbf{A} are lower for the uniform distribution than they are for the arcsine distribution (\fig{standardPD}\textbf{A}).\label{fig:standardPD_uniform}}
\end{figure}

\newpage

\begin{figure}
\centering
\includegraphics[width=0.9\textwidth]{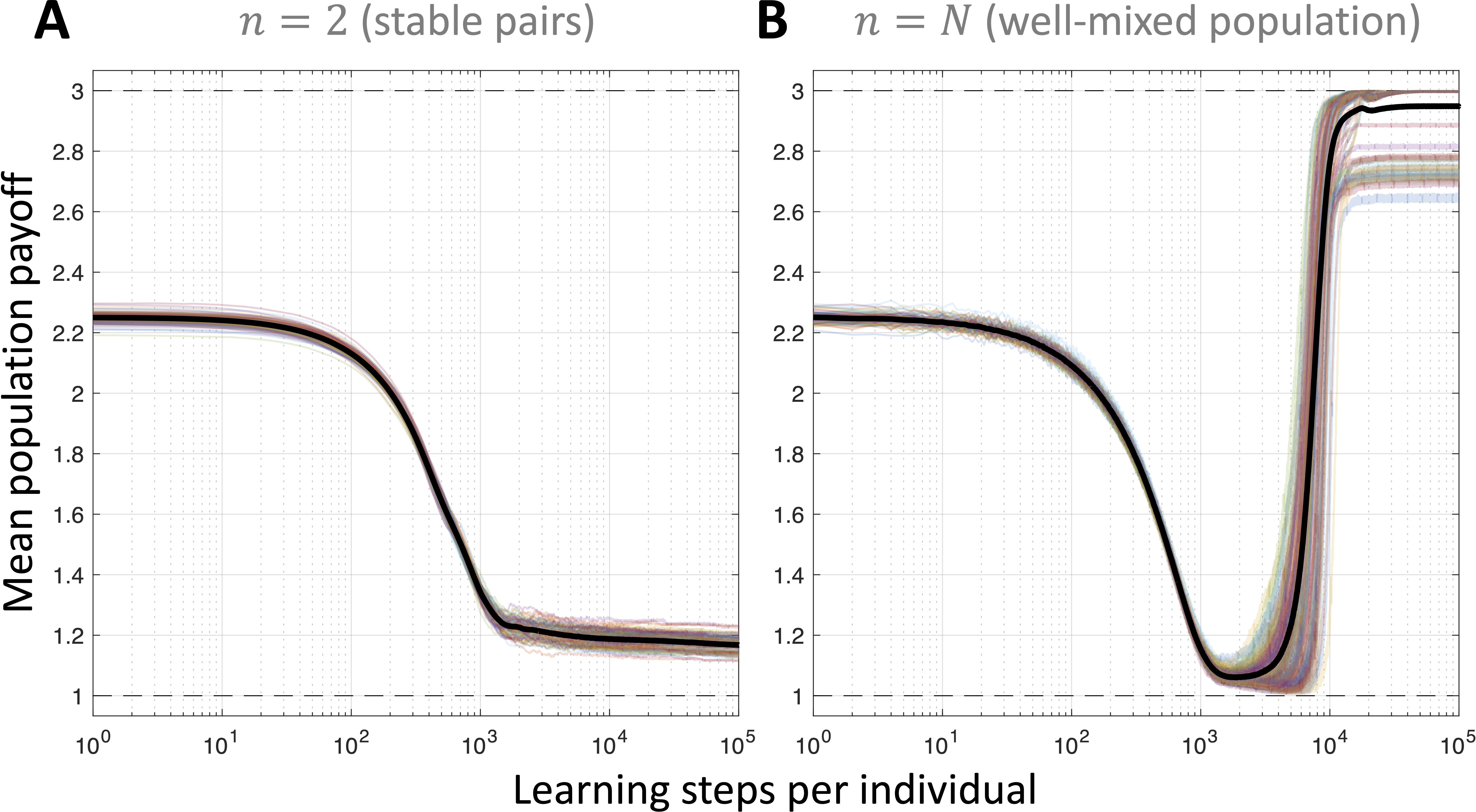}
\caption{\textbf{Pairwise versus collective learning in a standard prisoner's dilemma under random search.} Here, the setup is identical to that of \fig{standardPD}, with the exception of the optimization procedure, which is random search (as opposed to gradient ascent). In each encounter, learners are randomly paired and play a repeated game with one another. If $X$ uses $\p$ and interacts with $Y$, who uses $\q$, then $X$ samples $\p '$ by letting $p_{i}'$ be chosen uniformly from the interval $\left[p_{i}-s,p_{i}+s\right]$ (here, $s=10^{-2}$), truncating at $0$ and $1$, if necessary, by setting $p_{i}'=\min\left\{\max\left\{p_{i}',0\right\} ,1\right\}$. $X$ accepts this strategy if $\pi\left(\p ',\q\right) >\pi\left(\p ,\q\right) +\varepsilon$, where $\varepsilon$ (here, $\varepsilon =10^{-12}$) is a small threshold used to ensure that improvements in $X$'s payoff are detectable. Otherwise, $X$ retains the strategy $\p$. $Y$ simultaneously performs the same procedure, sampling $\q '$ and accepting it if and only if $\pi\left(\q ',\p\right) >\pi\left(\q ,\p\right) +\varepsilon$. The results are qualitatively similar to those of \fig{standardPD}, with outcomes in a well-mixed population much better than those of a population of stable pairs, although for this population size ($N=1{,}000$) some trajectories fall short of reaching the maximum possible payoff.\label{fig:standardPD_RS}}
\end{figure}

\newpage

\begin{figure}
\centering
\includegraphics[width=0.7\textwidth]{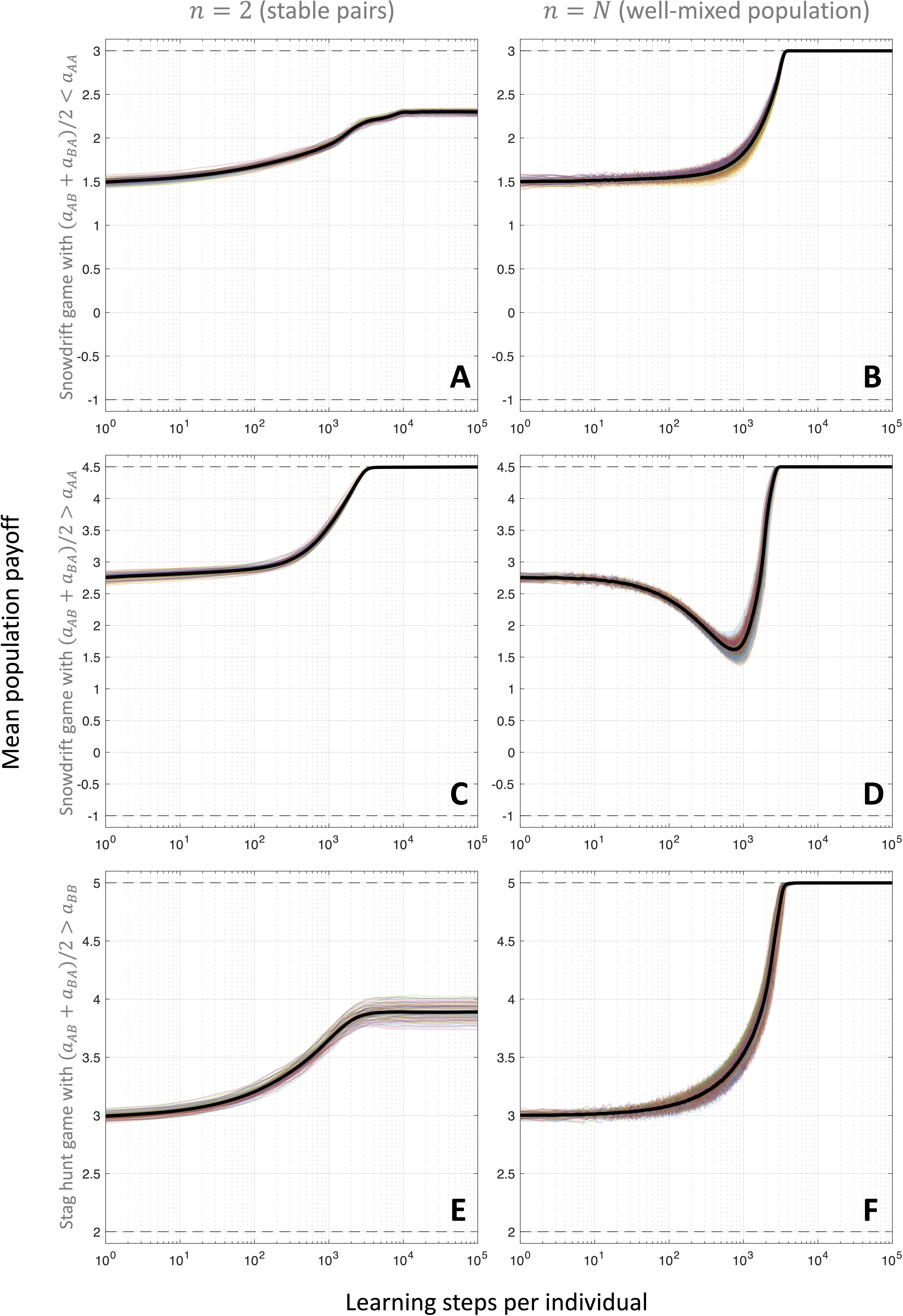}
\caption{\textbf{Selfish learning in other social dilemmas with $\left(a_{AB}+a_{BA}\right) /2>a_{BB}$.} Panels \textbf{A}--\textbf{D} depict snowdrift games, with $\left(a_{AA},a_{AB},a_{BA},a_{BB}\right) = \left(3,0,4,-1\right)$ in \textbf{A} and \textbf{B} and $\left(a_{AA},a_{AB},a_{BA},a_{BB}\right) = \left(3,0,9,-1\right)$ in \textbf{C} and \textbf{D}. Panels \textbf{E} and \textbf{F} show a stag hunt game, with parameters $\left(a_{AA},a_{AB},a_{BA},a_{BB}\right) = \left(5,1,4,2\right)$. In all panels, the population size is $N=1{,}000$, the number of runs is $100$, the continuation probability is $\lambda =0.9999$, and the learning rate is $\eta =10^{-3}$.\label{fig:GP}}
\end{figure}

\newpage

\begin{figure}
\centering
\includegraphics[width=0.8\textwidth]{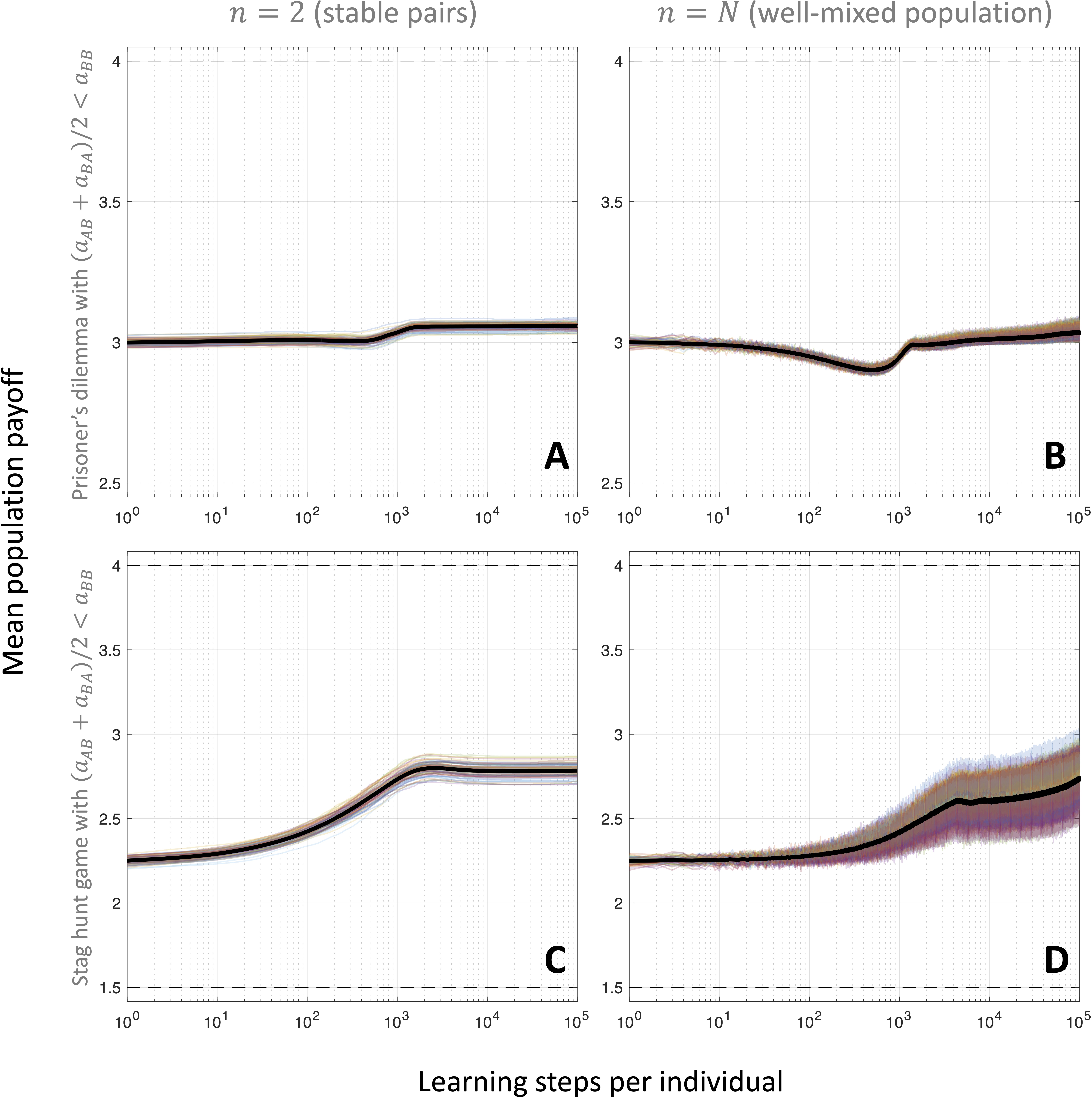}
\caption{\textbf{Selfish learning in social dilemmas with $\left(a_{AB}+a_{BA}\right) /2<a_{BB}$.} Panels \textbf{A} and \textbf{B} represent a prisoner's dilemma with $\left(a_{AA},a_{AB},a_{BA},a_{BB}\right) =\left(4,0,5,3\right)$. Panels \textbf{C} and \textbf{D}, depict a stag hunt game with $\left(a_{AA},a_{AB},a_{BA},a_{BB}\right) =\left(4,0,3,2\right)$. In all panels, the population size is $N=1{,}000$, the number of runs is $100$, the continuation probability is $\lambda =0.9999$, and the learning rate is $\eta =10^{-3}$.\label{fig:LP}}
\end{figure}

\newpage

\begin{figure}
\centering
\includegraphics[width=0.5\textwidth]{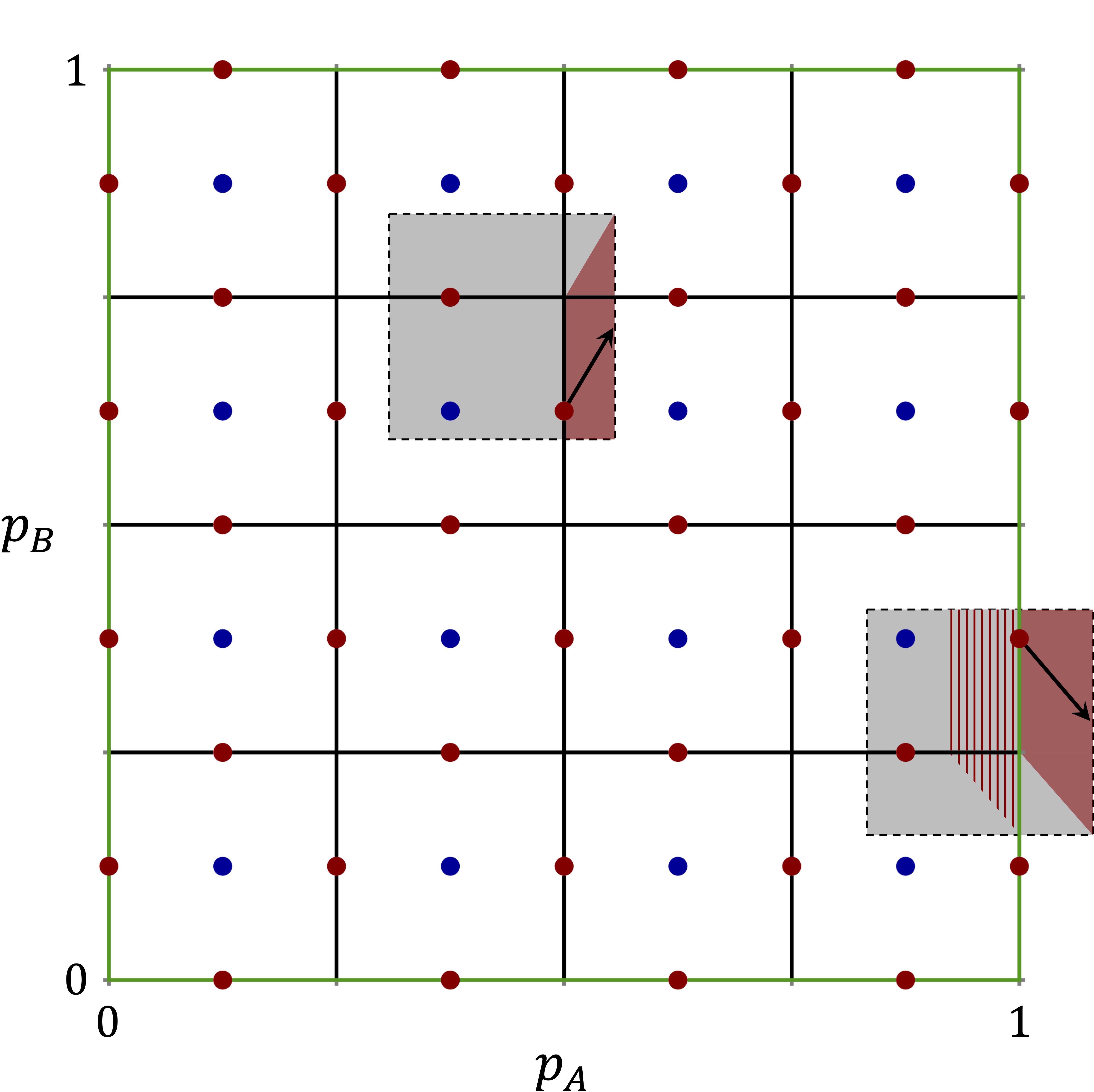}
\caption{\textbf{Illustration of the numerical scheme for the infinite-population limit.} In the limit $n=N\rightarrow\infty$, we track the evolution of strategy densities using the depicted numerical scheme. The space of reactive strategies is discretized by a grid, of box width $dp_{A}$ and height $dp_{B}$. Shown here is $dp_{A}=dp_{B}=1/4$; to generate \fig{PDE} in the main text, we use $dp_{A}=dp_{B}=1/50$. The density in each voxel is approximated by its value at the center (blue dots), and the flux (red regions, based on payoff gradients) is calculated for each of the four boundaries at the respective midpoints (red dots). At the boundary, we apply a translation (red, striped region) to be consistent with our use of gradient ascent since movement outside of the domain in the model is prevented by projecting gradient-based updates back into the unit cube.\label{fig:numerical_scheme}}
\end{figure}

\newpage

\renewcommand\figurename{Vid.}
\setcounter{figure}{0}

\begin{figure}
\centering
\caption{\textbf{Evolution of the probability density for reactive strategies.} For the numerical scheme in \fig{numerical_scheme}, this density, $\rho$, results in the mean payoff trajectories of \fig{PDE}. In the initial distribution, the coordinates of $\left(p_{A},p_{B}\right)$ are independent, each with an arcsine distribution. The dynamics of this density reveal a transition from random to one concentrated around unconditional defection ($\p =\left(0,0\right)$). Subsequently, strategies close to tit-for-tat ($\p =\left(1,0\right)$) emerge, which causes an exodus from unconditional defection, sending a wave out in the direction of the line $p_{B}=1$. Simultaneously, there is movement from tit-for-tat to more generous strategies, which still reciprocate cooperation but are also more likely to forgive defection. (Video available as an arXiv ancillary file.)\label{vid:density}}
\end{figure}

\end{document}